\documentclass[aps,prd,preprint,onecolumn,groupedaddress]{revtex4}

\usepackage{latexsym}
\usepackage{amsthm}
\usepackage{amsmath}
\usepackage{graphicx}
\usepackage{mathtools}
\usepackage{slashed}
\usepackage{amssymb}
\usepackage{indentfirst}
\usepackage{subfigure}
\usepackage{color}
\usepackage{epsf}
\definecolor{mycolor}{RGB}{0,30,100}
\usepackage[colorlinks=True, citecolor=blue, urlcolor=blue, linkcolor=blue]{hyperref}

\newcommand*\diff{\mathop{}\!\mathrm{d}}

\newcommand{\nn}{\nonumber}

\newcommand{\be}{\begin{eqnarray}}
\newcommand{\ee}{\end{eqnarray}}
\newcommand{\ma}{\mathrm}
\newcommand{\ml}{\mathcal}
\newcommand{\bs}{\boldsymbol}
\newcommand{\Tr}{\mathrm{Tr}}

\DeclareMathOperator{\sign}{sign}

\begin{document}

\title{Quarkonium In-Medium Transport Equation Derived from First Principles}

\author{Xiaojun Yao}
\email{xiaojun.yao@duke.edu}
\author{Thomas Mehen}
\email{mehen@phy.duke.edu}
%\author{Berndt M\"uller}
%\email{mueller@phy.duke.edu}
\affiliation{Department of Physics, Duke University, Durham, NC 27708, USA}

\date{\today}

\begin{abstract}
We use the open quantum system formalism to study the dynamical in-medium evolution of quarkonium. The system of quarkonium is described by potential non-relativistic QCD while the environment is a weakly coupled quark-gluon plasma in local thermal equilibrium below the melting temperature of the quarkonium. Under the Markovian approximation, it is shown that the Lindblad equation leads to a Boltzmann transport equation if a Wigner transform is applied to the system density matrix. Our derivation illuminates how the microscopic time-reversibility of QCD is consistent with the time-irreversible in-medium evolution of quarkonium states. Static screening, dissociation and recombination of quarkonium are treated in the same theoretical framework. In addition, quarkonium annihilation is included in a similar way, although the effect is negligible for the phenomenology of the current heavy ion collision experiments. The methods used here can be extended to study quarkonium dynamical evolution inside a strongly coupled QGP, a hot medium out of equilibrium or cold nuclear matter, which is important to studying quarkonium production in heavy ion, proton-ion, and electron-ion collisions.
\end{abstract}

\maketitle

\section{Introduction}
Heavy quarkonium production at hadron colliders has been studied extensively in both theory and experiment. In proton-proton collisions, the production process factorizes into a short-distance process of producing a heavy quark antiquark pair and a long-distance coalescence into a bound state \cite{Bodwin:1994jh}. In heavy ion collisions, the production process is complicated by the existence of a hot nuclear environment, the quark-gluon plasma (QGP). By comparing the quarkonium production in proton-proton and heavy ion collisions, one can study the properties of the hot medium produced during the collision (with the modification of the initial hard production due to heavy nuclei properly included). Static screening has been studied since the pioneering work of Ref.~\cite{Matsui:1986dk}, which provides a partial understanding of the the suppression of quarkonia in heavy ion collisions. For a more complete understanding, a theoretical description of quarkonium dynamics that also accounts for the dynamical screening and recombination inside the hot nuclear medium is needed.

There have been several approaches to address the question. First, statistical hadronization models have been used to describe charmonium production \cite{Andronic:2003zv,Andronic:2007bi}. In these models it is assumed that the charm quark evolves unbound inside the hot medium due to the Debye screening. During the evolution, the charm quark equilibrates kinematically but not chemically, because the total number of charm quarks is fixed by the initial hard scattering. The annihilation of charm quarks is negligible during the lifetime of the QGP. Thermal production is also negligible because of the large quark mass, compared with the medium temperature. Charmonium is produced from coalescence of charm quarks and antiquarks with thermal momenta at the transition hyper-surface of QGP to a hadron gas. Although the model has some phenomenological success, it is limited to the study of charmonium with low transverse momentum. The kinematic thermalization assumption is never justified for charmonium at large transverse momentum or for bottomonium.

Another approach is to use a transport equation \cite{Grandchamp:2003uw, Grandchamp:2005yw, Yan:2006ve, Liu:2009nb, Song:2011xi, Song:2011nu, Sharma:2012dy, Nendzig:2014qka, Krouppa:2015yoa, Chen:2017duy, Zhao:2017yan, Du:2017qkv, Aronson:2017ymv, Ferreiro:2018wbd}. In this approach, a rate equation is used to describe the dissociation and recombination of quarkonium inside the medium. Debye screening of the potential is also accounted for when solving the bound state wavefunction. In many studies, the dissociation rate is calculated from perturbative QCD while the recombination is modeled from detailed balance with an extra suppression factor accounting for the incomplete thermalization of heavy quarks. The recombination process has also been analyzed in the framework of perturbative QCD with parametrized non-thermal heavy quark momentum distributions \cite{Song:2012at}. Many studies have used potential non-relativistic QCD (pNRQCD) to study quarkonium dissociation rates inside the QGP \cite{Brambilla:2010vq, Brambilla:2011sg, Brambilla:2013dpa}. Recombination in a pNRQCD-based Boltzmann equation has been studied in Ref.~\cite{Yao:2017fuc}. New studies construct coupled Boltzmann transport equations of both heavy quarks and quarkonia, in which the heavy quark momentum distribution is not from an assumed parametrization but rather calculated from real-time dynamics, and quarkonium dissociation and recombination are calculated in the same theoretical framework \cite{Yao:2017fuc, Yao:2018zze, Yao:2018zrg}. By using the coupled Boltzmann transport equations, detailed balance and thermalization of heavy quark and quarkonium can be demonstrated from the real-time dynamics of heavy quark energy loss and the interplay between quarkonium dissociation and recombination.

More recently, an approach based on open quantum systems has been studied widely \cite{Young:2010jq, Borghini:2011ms, Akamatsu:2011se, Akamatsu:2014qsa, Blaizot:2015hya, Kajimoto:2017rel, DeBoni:2017ocl, Blaizot:2017ypk, Brambilla:2017zei, Akamatsu:2018xim}. In this approach, the system of the heavy quark and quarkonium and the medium evolve unitarily together. When the environment degrees of freedom are traced out, the system evolves non-unitarily and stochastic interactions can appear. This approach is a quantum description rather than a semi-classical equation. It has the advantage that non-unitarity appears automatically after tracing out the environment, while at the same time, preserving the total number of heavy quarks (by preserving the trace of the system density matrix). Quarkonium dissociation occurs during the non-unitary evolution but the unbound heavy quark pairs from dissociation never disappear from the system and they may recombine. This feature is never easily realized in transport models based on complex potentials. Another advantage is that the recombination effect is included systematically in this procedure. Meanwhile, the non-unitary time evolution is generally irreversible. For a general discussion of the occurrence of time-irreversible processes from time-reversible underlying theory, we refer to Ref.~\cite{Rau:1995ea}. The combination of the open quantum system and effective field theory (EFT) has also been recently used to study different physical systems: dissipative fluids \cite{Crossley:2015evo}, deep inelastic reactions \cite{Braaten:2016sja} and bottomonium suppression in Au-Au collisions \cite{Brambilla:2017zei}. 

In this paper, we demonstrate a deep connection between the approaches of open quantum systems and transport equations. More specifically, we use the open quantum system formalism, EFT of QCD, and the Wigner transform to derive the Boltzmann transport equation. Our derivation clarifies the conditions for the validity of quarkonium transport (rate) equations that are based on Boltzmann transport equations. We will justify the Markovian approximation in the open quantum system approach and the molecular chaos approximation in the Boltzmann equation. The work of Ref.~\cite{Brambilla:2017zei} focuses on quantum evolution of the density matrix and neglects center-of-mass (c.m.) motions of heavy quark anti-quark pairs. Thus, it is unable to study observables as functions of transverse momentum and rapidity of the quarkonium. In this work, we explicitly keep track of the c.m.~motion and focus on deriving the semi-classical Boltzmann transport equation from the quantum evolution of the system density matrix.

This paper is organized as follows: First, the open quantum system and the quantum master equation, the Lindblad equation, are briefly reviewed in Sec.~\ref{sect:lindblad}. The Boltzmann transport equation is derived in Sec.~\ref{sect:boltzmann}. Then quarkonium annihilation is studied similarly in Sec.~\ref{sect:annihilation}. Finally, conclusions are drawn in Sec.~\ref{sect:conclusion}.

\section{Lindblad equation in weakly-coupled system}
\label{sect:lindblad}
In this section we briefly review standard results in open quantum systems, which are covered in many textbooks, see, for example, Ref.~\cite{oqs_book1}. Assume the Hamiltonian of the system and environment (thermal bath) is given by
\be
H = H_S +H_B + H_I\,,
\ee
where $H_S$ is the system Hamiltonian, $H_B$ is the environment Hamiltonian, and $H_I$ contains the interactions between system and environment. The interaction Hamiltonian is assumed to be factorized as follows: $H_I = \sum_{\alpha} O^{(S)}_{\alpha} \otimes O^{(B)}_{\alpha}$ where $\alpha$ denotes all quantum numbers. The operators $O^{(S)}_{\alpha}$ are for the system while $O^{(B)}_{\alpha}$ are for the environment. We can assume $\langle O^{(B)}_{\alpha}\rangle \equiv \Tr_B(O^{(B)}_{\alpha}\rho_B) = 0$ because we can redefine $O^{(B)}_{\alpha}$ and $H_S$ by $O^{(B)}_{\alpha} - \langle O^{(B)}_{\alpha}\rangle$ and $H_S + \sum_{\alpha} O^{(S)}_{\alpha} \langle O^{(B)}_{\alpha}\rangle $ respectively. Here $\rho_B$ is the density matrix of the environment. Each part of the Hamiltonian is assumed to be Hermitian.

The von Neumann equation for the time evolution of the density matrix in the interaction picture is given by
\be
\frac{\diff \rho^{(\ma{int})}(t)}{\diff t} = -i [H^{(\ma{int})}_I(t), \rho^{(\ma{int})}(t)] \,.
\ee 
We will omit the superscript ``(int)" in the following. The symbolic solution is given by
\be
\rho(t) = U(t) \rho(0) U^{\dagger}(t)\,,
\ee
where the evolution operator is
\be
U(t) = \ml{T} e^{-i \int_0^t H_I(t') \diff t'}\,,
\ee
and $\ml{T}$ is the time-ordering operator.
We assume the interaction is a weak perturbation and expand the evolution operator to the second order in $H_I$:
\be
\label{eqn:2order}
\rho(t) &=& \rho(0) - i \int_0^t  \diff t' [H_I(t'), \rho(0) ] + \int_0^t \diff t_1 \int_0^t \diff t_2 \Big(H_I(t_1)\rho(0)H_I(t_2) \\ \nn
&& - \theta(t_1-t_2)H_I(t_1)H_I(t_2)\rho(0) - \theta(t_2-t_1)\rho(0)H_I(t_1)H_I(t_2)\  \Big) + \ml{O}(H_I^3)\,.
\ee
We shall assume the initial condition is given by
\be
\rho(0) = \rho_S(0) \otimes \rho_B\,,
\ee
where the environment density matrix is assumed to be time-independent. We define 
\be
\label{eqn:corr}
C_{\alpha\beta}(t_1, t_2) \equiv \Tr_B(O^{(B)}_{\alpha}(t_1)O^{(B)}_{\beta}(t_2)\rho_B)\,.
\ee
Then by taking the partial trace over the environment we can obtain the evolution equation of the system
\be\nn
&&\rho_S(t) = \Tr_B(\rho(t)) = \rho_S(0) - i\int_0^t\diff t'\sum_{\alpha} [O^{(S)}_{\alpha}(t'), \rho_S(0)] \Tr_B(O^{(B)}_{\alpha}(t')\rho_B) \\ \nn
&&  + \sum_{\alpha, \beta} \int_0^t \diff t_1 \int_0^t \diff t_2 C_{\alpha\beta}(t_1, t_2)  \Big(  O^{(S)}_{\beta}(t_2)\rho_S(0) O^{(S)}_{\alpha}(t_1) - \theta(t_1-t_2)O^{(S)}_{\alpha}(t_1)O^{(S)}_{\beta}(t_2)\rho_S(0) \\
&& - \theta(t_2-t_1)\rho_S(0) O^{(S)}_{\alpha}(t_1)O^{(S)}_{\beta}(t_2)    \Big) + \ml{O}(H_I^3)\,.
\ee
Using $\langle O^{(B)}_{\alpha}\rangle = 0$ and inserting complete sets of the system we obtain
\be
&&\rho_S(t) = \rho_S(0) + \sum_{\alpha, \beta} \int_0^t \diff t_1 \int_0^t \diff t_2 C_{\alpha\beta}(t_1, t_2) \sum_{a,b,c,d} \langle a | O^{(S)}_{\beta}(t_2) | b\rangle \langle c | O^{(S)}_{\alpha}(t_1) | d \rangle ^* \\ \nn
&&\Big( |a\rangle\langle b| \rho_S(0) ( |c\rangle\langle d|)^{\dagger}  -\theta(t_1-t_2) ( |c\rangle\langle d|)^{\dagger}|a\rangle\langle b|\rho_S(0) -  \theta(t_2-t_1) \rho_S(0) ( |c\rangle\langle d|)^{\dagger}|a\rangle\langle b| \Big) + \ml{O}(H_I^3)\,.
\ee
Finally defining the Lindblad operator $L_{ab} \equiv |a\rangle \langle b|$ and
\be
\gamma_{ab,cd} (t) &\equiv& \sum_{\alpha, \beta} \int_0^t \diff t_1 \int_0^t \diff t_2 C_{\alpha\beta}(t_1, t_2) \langle a | O^{(S)}_{\beta}(t_2) | b\rangle \langle c | O^{(S)}_{\alpha}(t_1) | d \rangle ^* \\
\sigma_{ab}(t) &\equiv& \frac{-i}{2} \sum_{\alpha, \beta} \int_0^t \diff t_1 \int_0^t \diff t_2 C_{\alpha\beta}(t_1, t_2) \sign(t_1-t_2) \langle a | O^{(S)}_{\alpha}(t_1) O^{(S)}_{\beta}(t_2) | b\rangle \,,
\ee
we obtain the Lindblad equation up to second order in perturbation theory
\be \nn
\rho_S(t) = \rho_S(0) + \sum_{a,b,c,d} \gamma_{ab,cd} (t) \Big( L_{ab}\rho_S(0)L^{\dagger}_{cd} - \frac{1}{2}\{ L^{\dagger}_{cd}L_{ab}, \rho_S(0)\}  \Big)\\
\label{eqn:lindblad}
 -i\sum_{a,b} \sigma_{ab}(t) [L_{ab}, \rho_S(0)] + \ml{O}(H_I^3)\,.
\ee
The relation $\theta(t)=(1+\sign(t))/2$ has been used in the derivation. It will be shown in the next section that for quarkonium, the commutator term is a loop correction of the real part of the Hamiltonian. The anticommutator term describes the dissociation of quarkonium, which can also be thought of as an imaginary part of the potential. The second term on the right hand side of Eq.~(\ref{eqn:lindblad}) represents the recombination contribution.  A direct conclusion from Eq.~(\ref{eqn:lindblad}) is the conservation of probability: $\Tr{\rho_S(t)} = \Tr{\rho_S(0)} $. This implies the unbound heavy quark antiquark pair from quarkonium dissociation stays as active degrees of freedom of the system and may recombine later in the evolution.

The form of the Lindblad equation is valid up to all orders in the perturbative expansion \cite{Breuer:2002pc}. So the higher-order terms neglected here can also be written in the form of the Lindblad equation. The Lindblad equation cannot be written in the form of a von Neumann equation because the evolution is non-unitary. The time-irreversibility can be seen by noting that the relative entropy of the system with respect to a steady state under the partial trace is monotonically decreasing \cite{Breuer:2002pc}. The partial trace over the environment can be thought of as an average over different environment configurations. Though the dynamics involving each configuration is governed by a time-reversible theory with a unitary evolution, after averaging, the dynamics becomes time-irreversible and non-unitary.

\section{Derivation of Boltzmann equation}
\label{sect:boltzmann}
In this section we will derive the Boltzmann transport equation by applying the Lindblad equation, Eq.~(\ref{eqn:lindblad}), to the Wigner transform of the density matrix describing heavy quark antiquark pairs that can be bound or unbound. The system in vacuum can be described by pNRQCD \cite{Brambilla:1999xf, Fleming:2005pd}. The effective theory can be constructed from QCD by a non-relativistic expansion assuming the separation of scales: $M\gg Mv\gg Mv^2$, where $M$ is the heavy quark mass and $v$ is the velocity of the heavy quark antiquark inside a quarkonium. The quarkonium size is roughly given by $r\sim 1/(Mv)$. The environment is a weakly-coupled QGP in local thermal equilibrium, $\rho_B = \frac{1}{\ml{Z}}e^{-\beta H_B}$, where $\ml{Z} = \Tr_Be^{-\beta H_B}$. Thus the correlations in Eq.~(\ref{eqn:corr}) can be calculated in real-time thermal field theory. We review different definitions of thermal correlations (Green's functions) in Appendix~\ref{app:real-time}. We will use free thermal Green's functions of gauge fields. Our derivation can be extended by using resummed thermal propagators. Resummed thermal propagators and pNRQCD have been used to investigate static heavy quark antiquark pairs at finite temperature \cite{Brambilla:2008cx}. The plasma provides two extra scales: the temperature $T$ and the Debye mass $m_D$ (we use units in which $k_B=1$). Here we will focus on the case where quarkonium exists as a well-defined bound state in a QGP that is below the melting temperature of the quarkonium, so  $M\gg Mv\gg Mv^2 \gtrsim T \gtrsim m_D$. We do not consider cases with $Mv\gg T \gg Mv^2$ or $Mv\sim T \gg Mv^2$ because $Mv^2 \sim 500$ MeV for charmonium and bottomonium, and the temperatures realized in current heavy ion experiments are smaller than this. For our choice of scaling both dissociation and recombination are possible.

PNRQCD can be constructed by matching with NRQCD at the scale $Mv$. The matching can be done perturbatively if $Mv\gg\Lambda_{QCD}$ or non-perturbatively. In either case, the quarkonium interacts with gluons from the QGP via a dipole interaction at lowest order. As will be seen below, the dipole interaction scales as $rT\sim\frac{T}{Mv} \lesssim v$, which is small in the assumed separation of scales. We assume a perturbative matching throughout the paper. The dipole interaction is not running at one-loop level \cite{Pineda:2000gza, Yao:2018sgn}, which means the coupling constant in the dipole term is set at the scale of $Mv$, no matter the scale of the scattering. To make calculations easier here, we follow Ref.~\cite{Fleming:2005pd} and use a slightly different notation for the pNRQCD Lagrangian density.
\be
\label{eq:lagr}\nn
\ml{L}_\ma{pNRQCD}({\bs R}, t) &=&  \ml{L}_{\ma{kin},s} + \ml{L}_{\ma{kin},o} + \ml{L}_{\ma{int},so} + \ml{L}_{\ma{int},oo} +\cdots \\ \nn
\ml{L}_{\ma{kin},s}  &=&  \langle S(\bs R, t) | (i\partial_0-H_s) | S(\bs R, t)\rangle \\ \nn
\ml{L}_{\ma{kin},o}  &=&  \langle O^a(\bs R, t) | ( i\partial_0-H_o )  |O^a(\bs R, t)\rangle \\ \nn
\label{eqn:so_interaction}
\ml{L}_{\ma{int},so} &=&  \sqrt{\frac{T_F}{N_C}}\Big( \langle O^a(\bs R, t) | \bs r \cdot g{\bs E}^a(\bs R, t) | S(\bs R, t)\rangle + \ma{h.c.} \Big) \\ \nn
\label{eqn:oo_interaction}
\ml{L}_{\ma{int},oo} &=& if^{abc} \langle O^a(\bs R, t) | g A_0^b(\bs R, t)  |O^c(\bs R, t)\rangle \\
&&  + d^{abc}  \langle O^{a}(\bs R, t)| g \bs r \cdot \bs E^b(\bs R, t) | O^{c}(\bs R, t)\rangle +\cdots  \, .
\ee
The degrees of freedom, in the standard pNRQCD Lagrangian, are the color singlet $S({\bs R}, {\bs r}, t)$ and octet $O^a({\bs R}, {\bs r}, t)$ where ${\bs R}$ and ${\bs r}$ are the center-of-mass (c.m.) and relative positions of the heavy quark antiquark pair. Here we define the ``bra-ket" notation via
\be \nn
\langle {\bs r} | S({\bs R}, t) \rangle &\equiv& S({\bs R}, {\bs r}, t) \\ \nn
\langle {\bs r} | O^a({\bs R}, t) \rangle  &\equiv& O^a({\bs R}, {\bs r}, t) \\
\langle S({\bs R}, t) | f({\bs r}) | O^a({\bs R}, t) \rangle &\equiv& \int\diff^3r S^\dagger({\bs R}, {\bs r}, t) f({\bs r}) O^a({\bs R}, {\bs r}, t)\,,
\ee
for any function $f$ of ${\bs r}$.
We use the ``bra-ket" notation so that we no longer need to write the integral over ${\bs r}$ explicitly, which simplifies notations in the derivation.
Summations over color indexes are assumed and higher-order terms in the velocity expansion are neglected. Here $N_c=3$, $T_F=\frac{1}{2}$. We define $C_F\equiv\frac{T_F}{N_c}(N_c^2-1)$ for later use.  The covariant derivative on the octet field has been written out explicitly $D_0O=\partial_0-ig[A_0, O]$. The Hamiltonians are expanded in powers of $v^2$,
\be
H_{s,o} = \frac{\bs p_\ma{cm}^2}{4M} + \frac{\bs p_\ma{rel}^2}{M} + V_{s,o}^{(0)} + \frac{V_{s,o}^{(1)}}{M} + \frac{V_{s,o}^{(2)}}{M^2} + \cdots\,.
\ee
We will work up to the leading order (LO) in $v^2$ and $\frac{\bs p_\ma{rel}^2}{M} \sim V_{s,o}^{(0)}\sim M v^2$ by the virial theorem. When the medium is static in the rest frame of quarkonium, the quarkonium exchanges gluons with the medium whose momentum and energy are $\sim T$ and gains a c.m.~momentum $\sim T$. In our power counting, $T\lesssim Mv^2$ and hence the c.m.~kinetic energy, $\frac{\bs p_\ma{cm}^2}{4M}$, is $\ml{O}(Mv^4)$ and is therefore neglected. If the medium moves with respect to the quarkonium at a velocity $v_\ma{med}$, the gluon energy is boosted to be $\frac{T}{\sqrt{1-v_\ma{med}^2}}$. The c.m.~kinetic energy is still suppressed if $v_\ma{med}\lesssim \sqrt{1-v}$. We assume $v_\ma{med}=0$ in the following but generalization to $v_\ma{med}\neq 0$ can be easily done by boosting the gluon distribution function. We do keep track of the c.m.~momentum so that momentum is conserved. The singlet and octet composite fields are given by
\be \nn
|S(\bs R, t) \rangle &=& \int\frac{\diff^3 p_\ma{cm}}{(2\pi)^3}  e^{-i(Et-\bs p_\ma{cm} \cdot \bs R)} \bigg( \sum_{nl} a_{nl}(\bs p_\ma{cm}) \otimes | \psi_{nl} \rangle   + \int\frac{\diff^3 p_{\ma{rel}}}{(2\pi)^3} b_{{\bs p}_{\ma{rel}}}(\bs p_\ma{cm}) \otimes | \psi_{{\bs p}_{\ma{rel}}} \rangle \bigg) \\
|O^a(\bs R, t) \rangle &=&  \int\frac{\diff^3 p_\ma{cm}}{(2\pi)^3} e^{-i(Et-\bs p_\ma{cm}\cdot \bs R)}  \int\frac{\diff^3 p_{\ma{rel}}}{(2\pi)^3} c^a_{{\bs p}_{\ma{rel}}}(\bs p_\ma{cm}) 
\otimes | \Psi_{{\bs p}_{\ma{rel}}} \rangle \,,
\ee
where $E$ is the eigenenergy of a state in the whole Hilbert space. The whole Hilbert space factorizes into two parts: one part for the c.m.~motion and the other for the relative motion. The operators $a^{(\dagger)}_{nl}(\bs p_\ma{cm})$, $b^{(\dagger)}_{{\bs p}_{\ma{rel}}}(\bs p_\ma{cm})$ and $c^{a(\dagger)}_{{\bs p}_{\ma{rel}}}(\bs p_\ma{cm})$ act on the Fock space to annihilate (create) composite particles with the c.m.~momentum ${\bs p_\ma{cm}}$ and the corresponding quantum numbers in the relative motion. These quantum numbers can be $nl$ for a bound singlet state, ${\bs p}_{\ma{rel}}$ for an unbound singlet state and color $a$ and ${\bs p}_{\ma{rel}}$ for an unbound octet state. When we compute the square of matrix elements, we will average over the polarizations of non-$S$ wave quarkonium states. In our notation, we omit the quantum number $m$ of the bound singlet state. In the octet channel no bound state exists because of the repulsive octet potential. The corresponding wavefunctions of the relative motion are $| \psi_{nl} \rangle$, $| \psi_{{\bs p}_{\ma{rel}}} \rangle$ and $ | \Psi_{{\bs p}_{\ma{rel}}} \rangle $. They can be obtained by solving the equations of motion of the free composite fields, which are Schr\"odinger equations. The eigenenergies are $E = -|E_{nl}|$ and $E = \frac{{\bs p}_{\ma{rel}}^2}{M}$ for the bound and unbound states separately with higher-order terms in $v$ neglected. Here $E_{nl}$ is the binding energy of the bound state $| \psi_{nl} \rangle$. The annihilation and creation operators in the Fock space satisfy the following commutation relations:
\be \nn
[a_{n_1l_1}({\bs p}_\ma{cm1}),\ a^{\dagger}_{n_2l_2}({\bs p}_\ma{cm2})] &=& (2\pi)^3 \delta^3({\bs p}_\ma{cm1} - {\bs p}_\ma{cm2}) \delta_{n_1n_2}\delta_{l_1l_2} \\ \nn
{[b_{{\bs p}_{\ma{rel}1}}({\bs p}_\ma{cm1}),\ b^{\dagger}_{{\bs p}_{\ma{rel}2}}({\bs p}_\ma{cm2})]} & =& (2\pi)^6 \delta^3({\bs p}_\ma{cm1} - {\bs p}_\ma{cm2})\delta^3({\bs p}_{\ma{rel}1}-{\bs p}_{\ma{rel}2}) \\
{[c^{a_1}_{{\bs p}_{\ma{rel}1}}({\bs p}_\ma{cm1}),\ c^{a_2\dagger}_{{\bs p}_{\ma{rel}2}}({\bs p}_\ma{cm2})]} &=& (2\pi)^6 \delta^3({\bs p}_\ma{cm1} - {\bs p}_\ma{cm2}) \delta^3({\bs p}_{\ma{rel}1}-{\bs p}_{\ma{rel}2}) \delta^{a_1a_2}\,,
\ee
and all other commutators vanish. 

The interaction part of the Hamiltonian of the theory is given in Eq.~(\ref{eqn:so_interaction}) but only the singlet-octet transition is relevant for the dissociation and recombination of quarkonium. The octet-octet interaction governs the dynamical evolution of unbound heavy quarks and thus is only present in the transport equation of open heavy quarks. We will neglect the octet-octet interaction when deriving the quarkonium transport equation. The minus sign in the Hamiltonian is of no importance at the order $\ml{O}(H_I^2)$. The weak coupling expansion in $H_I$ is valid because the quarkonium size is small $rT \sim \frac{T}{Mv} \lesssim v$ in our power counting. For current heavy ion experiments, this assumption should hold for the most compact quarkonia such as the $\Upsilon$(1S). It could work as well for the $\Upsilon$(2S) if the temperature is below $\sim$200 MeV. As discussed above, this is true in both perturbative and non-perturbative constructions of the pNRQCD. When $rT\sim1$, the static screening effect of the potential is too strong to support the quarkonium bound state.

To use the Lindblad equation derived in Sec~\ref{sect:lindblad}, we write $H_I$ as $\sum_{\alpha} O^{(S)}_{\alpha} \otimes O^{(B)}_{\alpha}$ with
\be \nn
O^{(S)}_{\alpha} &\rightarrow& \langle S(\bs R, t) | r_i | O^a(\bs R, t)\rangle  + \langle O^{a}(\bs R, t) | r_i | S(\bs R, t)\rangle \\
O^{(B)}_{\alpha} &\rightarrow& \sqrt{\frac{T_F}{N_C}}g E_i^{a}(\bs R, t)\,.
\ee
The sum over $\alpha$ means
\be
\sum_{\alpha} \rightarrow \int \diff^3 R \sum_i \sum_a \,.
\ee
The complete set used to construct the Lindblad operators are
\be \nn
|\bs k, nl, 1\rangle &=& a^{\dagger}_{nl}(\bs k) | 0 \rangle \\ \nn
 |{\bs p}_{\ma{cm}}, {\bs p}_{\ma{rel}} ,1\rangle  &=& b^{\dagger}_{{\bs p}_{\ma{rel}}}({\bs p}_{\ma{cm}}) | 0 \rangle\\
 |{\bs p}_{\ma{cm}}, {\bs p}_{\ma{rel}}, a\rangle  &=& c^{a\dagger}_{{\bs p}_{\ma{rel}}}({\bs p}_{\ma{cm}})  | 0 \rangle\,,
\ee
where $1$ denotes the singlet while $a$ is the color index of an octet. The unbound singlet state will not be used in our current calculation because at the order we are working, an unbound singlet cannot form a bound singlet by radiating out one gluon; only an unbound octet can do so.

We are interested in the bound state evolution. Therefore our basic strategy is to study the time evolution of $\langle {\bs k}_1, n_1l_1, 1 | \rho_S(t) | {\bs k}_2, n_2l_2, 1\rangle $ by sandwiching Eq.~(\ref{eqn:lindblad}) between $\langle {\bs k}_1, n_1l_1, 1 |$ and $ | {\bs k}_2, n_2l_2, 1\rangle $. To obtain the evolution equation of the semi-classical phase space distribution function, we will take the Wigner transform of the density matrix
\be
\label{eqn:wigner}
f_{nl}({\bs x}, {\bs k}, t) \equiv \int\frac{\diff^3k'}{(2\pi)^3} e^{i {\bs k}'\cdot {\bs x} } \langle  {\bs k}+\frac{{\bs k}'}{2}, nl,1   | \rho_S(t)  |   {\bs k}-\frac{{\bs k}'}{2} , nl, 1\rangle \,.
\ee
We will extract the linear dependence on $t$ of $\gamma_{ab,cd}$ and $\sigma_{ab}$ terms in Eq.~(\ref{eqn:lindblad}) and then take time derivative at $t=0$ on both sides of Eq.~(\ref{eqn:lindblad}). The double time integrals are simplified by assuming the Markovian approximation, i.e., the upper limit of the time integrals is large and can be taken to be infinity, $t\rightarrow\infty$. The Markovian approximation is valid when the environment correlation time is much smaller than the relaxation time of the system. The former is roughly given by $1/T$ while the latter can be estimated by the inverse of the dissociation rate. The dissociation rate is $\sim(grT)^2T\lesssim \alpha_s v^2T$ in our power counting and $\alpha_s$ is at the scale $Mv$. So in the assumed separation of scale, the Markovian approxiamtion is valid. The $t\rightarrow0$ limit in the time derivative and the $t\rightarrow\infty$ limit in the integral are not contradictory because the timescale of measuring the macroscopic phase space distribution, given by the Wigner transform of the density matrix, is much larger than the timescale of the microscopic dynamics. The Markovian approximation means that there is no memory effect \cite{Breuer:2002pc}. The absence of memory effect is reflected in the Boltzmann equation in the assumption of molecular chaos, namely that the correlation between particles generated from their previous collisions is completely forgotten in the next collision \cite{DavidTong:note}. Under the assumption of $t\rightarrow\infty$, the double time integrals give two delta functions in energy. When the two delta functions correspond to the same energy conservation, one can write them as one delta function multiplied by the time length $t$. This is how we extract the linear dependence in $t$. This trick is also used in the derivation of Fermi's golden rule.
Details of the derivation can be found in the Appendixes~\ref{app:static}, \ref{app:dissociation} and \ref{app:recombination}.

\begin{figure}
\centering
\includegraphics[width=0.6\linewidth]{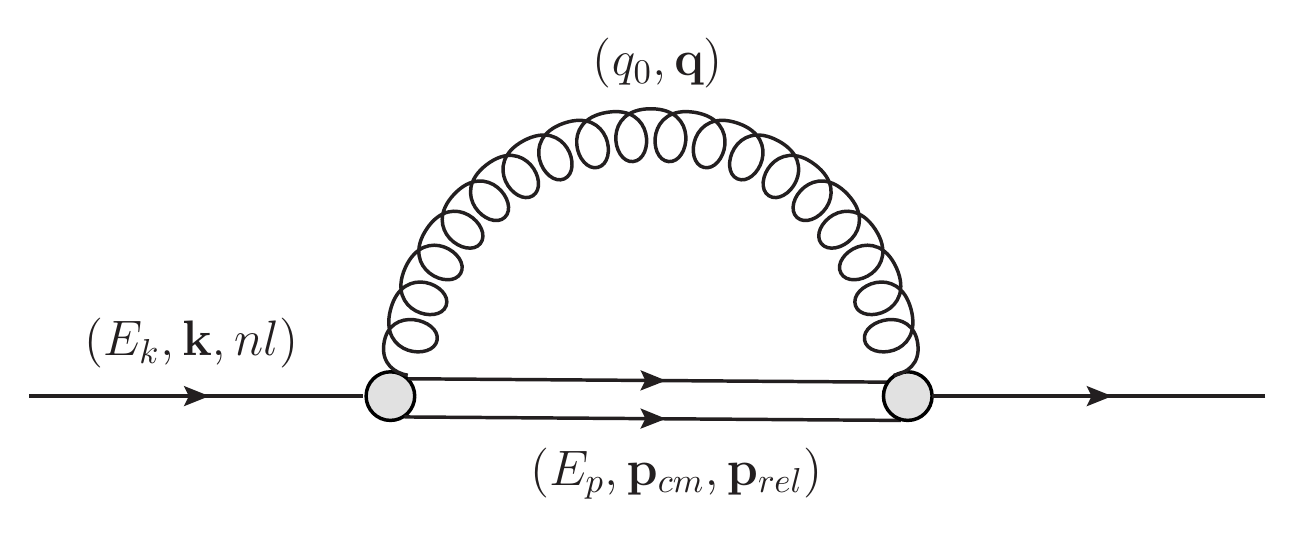}
\caption{Loop correction (self-energy) of the singlet field. A single solid line indicates the bound singlet state while the double solid lines represent the unbound octet state.}
\label{fig:loop_singlet}
\end{figure}

First, the $\sum_{a,b}\sigma_{ab}L_{ab}$ term in the Lindblad equation,  Eq.~(\ref{eqn:lindblad}), can be shown to give, for the bound singlet part
\be \nn
\sum_{a,b}\sigma_{ab}L_{ab} &\rightarrow& t \sum_{n,l} \int\frac{\diff^3k}{(2\pi)^3} \Re  \bigg\{ -i g^2 C_F \sum_{i_1,i_2} \int\frac{\diff^4q}{(2\pi)^4} 
\int\frac{\diff^4 p_\ma{cm}}{(2\pi)^4} \int\frac{\diff^3 p_\ma{rel}}{(2\pi)^3}  \\ \nn
&& (2\pi)^4\delta^3 ({\bs k}-{\bs p}_\ma{cm} - {\bs q}) \delta(E_{k}-p_\ma{cm}^0-q^0) \\\nn
&& (q_0^2 \delta_{i_1i_2} -q_{i_1}q_{i_2}) \Big( \frac{i}{q_0^2 - {\bs q}^2 + i\epsilon} +   n_B(|q_0|)(2\pi)  \delta(q_0^2-{\bs q}^2)  \Big) \\ 
&& \langle \psi_{nl} | r_{i_1}  \frac{i  | \Psi_{{\bs p}_\ma{rel}} \rangle  \langle \Psi_{{\bs p}_\ma{rel}} |}{p_{\ma{cm}}^0-E_{p}+i\epsilon}   r_{i_2} |  \psi_{nl} \rangle 
\bigg\} L_{| {\bs k},nl,1 \rangle \langle  {\bs k},nl,1   |} \,.
\ee
The part inside the curly brackets gives the loop correction of the potential, which can be calculated as usual by the standard quantum field theory perturbative technique: computing the loop shown in Fig.~\ref{fig:loop_singlet} by using the time-ordered propagators. Only the real part of the correction contributes here.

This correction $\sum_{a,b}\sigma_{ab}L_{ab}$ is diagonal in the bound state space and is Hermitian. For our purpose, we may write $\sum_{a,b}\sigma_{ab}L_{ab}\equiv t H_{\ma{1-loop}}$. Recall that if we go back to the Schr\"odinger picture
\be
\rho_S(t) = \rho_S(0)-it [H_S, \rho_S(0)] - i [\sum_{a,b}\sigma_{ab}L_{ab},  \rho_S(0)] + \cdots
\ee
where other terms in the Lindblad Eq.~(\ref{eqn:lindblad}) have been omitted temporarily. Now we define an effective Hamiltonian $H_\ma{eff} = H_S + H_{\ma{1-loop}} $. If the correction is perturbative, we can start with a potential in $H_S$, calculate wavefunctions of relative motions and the one-loop correction to the real part of the potential to obtain $H_\ma{eff} $. In some cases, it may be necessary to resum all the loop corrections of the real part of the potential into $H_\ma{eff}$ and then use $H_\ma{eff}$ to calculate the wavefunction of the relative motion. In this case, the real part of the potential in $H_\ma{eff}$ can be modeled by using recent high statistics lattice studies of the color singlet free energy at finite temperature \cite{Bazavov:2018wmo}. In this work, the explicit forms of the wavefunction are not needed.

Now if we do a Wigner transform of the form Eq.~(\ref{eqn:wigner}) on
\be
\rho_S(t) = \rho_S(0) -it(H_\ma{eff}\rho_S(0) - \rho_S(0)H_\ma{eff}) + \cdots \,,
\ee
we obtain 
\be 
f_{nl}({\bs x}, {\bs k}, t) 
%&=& -it\int\frac{\diff^3k'}{(2\pi)^3} e^{i {\bs k}'\cdot \bs x} \langle \bs k+\frac{{\bs k}'}{2}, nl, 1| (H_{eff}\rho_S -\rho_SH_{eff}) | \bs k-\frac{{\bs k}'}{2}, nl, 1 \rangle + \cdots\\
&=&f_{nl}({\bs x}, {\bs k}, 0)  \\ \nn
&-& it\int\frac{\diff^3k'}{(2\pi)^3} e^{i {\bs k}'\cdot \bs x} (E_{\bs k +\frac{{\bs k}'}{2}} - E_{\bs k -\frac{{\bs k}'}{2}}) \langle \bs k+\frac{{\bs k}'}{2}, nl, 1| \rho_S(0) | \bs k-\frac{{\bs k}'}{2}, nl, 1 \rangle + \cdots\,.
\ee
Here if we restore the c.m.~kinetic energy,
\be
E_{\bs k \pm \frac{{\bs k}'}{2}}  = -|E_{nl}| + \frac{(\bs k \pm \frac{{\bs k}'}{2})^2}{4M}\,,
\ee
we can write
\be \nn
f_{nl}({\bs x}, {\bs k}, t) &=& f_{nl}({\bs x}, {\bs k}, 0)  -it\int\frac{\diff^3k'}{(2\pi)^3} \frac{\bs k}{2Mi}\cdot\nabla_{\bs x}e^{i {\bs k}'\cdot \bs x}  \langle \bs k+\frac{{\bs k}'}{2}, nl, 1| \rho_S(0) | \bs k-\frac{{\bs k}'}{2}, nl, 1 \rangle + \cdots\\
\label{eqn:stream_f}
&=& f_{nl}({\bs x}, {\bs k}, 0)  - t {\bs v} \cdot\nabla_{\bs x} f_{nl}(\bs x, \bs k, 0) + \cdots \,,
\ee
where the c.m.~velocity of the quarkonium is defined as $\bs v = \frac{\bs k}{2M}$. 

Now we proceed to compute the contributions from the other two terms in the Lindblad Eq.~(\ref{eqn:lindblad}) omitted in Eq.~(\ref{eqn:stream_f}). The $-\sum_{a,b,c,d}\frac{1}{2}\gamma_{ab,cd}\{ L_{cd}^\dagger L_{ab}, \rho_S(0)\}$ term gives
\be \nn
 &&-t\int\frac{\diff^3p_{\ma{cm}}}{(2\pi)^3} \frac{\diff^3p_{\ma{rel}}}{(2\pi)^3} \frac{\diff^3q}{(2\pi)^32q} n_B(q) (2\pi)^4\delta^3({\bs k} - {\bs p}_{\ma{cm}} +{\bs q} ) \delta(-|E_{nl}|+q-\frac{{\bs p}_{\ma{rel}}^2}{M}) \\
 \label{eqn:disso_f}
 &&  \frac{2}{3}C_Fq^2g^2 | \langle \psi_{nl} | \bs r | \Psi_{{\bs p}_{\ma{rel}}} \rangle|^2   f_{nl}(\bs x, \bs k ,t=0) \equiv -t \ml{C}_{nl}^{(-)}({\bs x}, {\bs k}, t=0) \,.
\ee
The dissociation rate $\Gamma_{\ma{disso}}$ of a quarkonium with momentum ${\bs k}$ and position ${\bs x}$ can be defined by $\Gamma_{\ma{disso}}(\bs x, \bs k ,t) = \frac{\ml{C}_{nl}^{(-)}({\bs x}, {\bs k}, t) }{f_{nl}(\bs x, \bs k ,t)}$. The dissociation rate derived here is the same as calculated in Ref.~\cite{Brambilla:2011sg} by taking the imaginary part of Fig.~\ref{fig:loop_singlet}. The same dissociation term has been used in the Boltzmann transport equation in Ref.~\cite{Yao:2017fuc}.

The $\sum_{a,b,c,d}\gamma_{ab,cd} L_{ab} \rho_S(0) L_{cd}^\dagger $ term gives
\be
\label{eqn:reco_f}
&&t \sum_{a,i} \int\frac{\diff^3 p_{\ma{cm}}}{(2\pi)^3} \frac{\diff^3 p_{\ma{rel}}}{(2\pi)^3} \frac{\diff^3q}{(2\pi)^32q} (1+n_B(q))(2\pi)^4\delta^3({\bs k}-{\bs p}_{\ma{cm}}+\bs q) \delta(-|E_{nl}|+q-\frac{{\bs p}_{\ma{rel}}^2}{M})\\ \nn
&& \frac{2T_F}{3N_C}q^2g^2  \langle  \psi_{nl}  | r_i |  \Psi_{{\bs p}_{\ma{rel}}}  \rangle \int \diff^3r \,  \psi_{nl}(\bs r) r_i \Psi^*_{{\bs p}_{\ma{rel}}}(\bs r)
f_{Q\bar{Q}}(\bs x, {\bs p}_{\ma{cm}}, {\bs r}, {\bs p}_{\ma{rel}}, a,t=0) \equiv t\ml{C}_{nl}^{(+)}({\bs x}, {\bs k}, 0)\,,
\ee
where $f_{Q\bar{Q}}(\bs x, {\bs p}_{\ma{cm}}, {\bs r}, {\bs p}_{\ma{rel}}, a,t=0)$ is the two-particle distribution function of a heavy quark antiquark pair in color octet $a$ with the c.m.~position $\bs x$ and momentum ${\bs p}_{\ma{cm}}$ and relative position $\bs r$ and momentum ${\bs p}_{\ma{rel}}$. Unlike in the dissociation term, one of the integrals over the wavefunctions of the relative motion involves the two-particle distribution function of $Q\bar{Q}$.

Now putting Eqs.~(\ref{eqn:stream_f}), (\ref{eqn:disso_f}) and (\ref{eqn:reco_f}) together we finally infer the Boltzmann transport equations
\be
\frac{\partial}{\partial t} f_{nl}({\bs x}, {\bs k}, t) + {\bs v}\cdot \nabla_{\bs x}f_{nl}({\bs x}, {\bs k}, t) = \ml{C}_{nl}^{(+)}({\bs x}, {\bs k}, t) - \ml{C}_{nl}^{(-)}({\bs x}, {\bs k}, t)\,
\ee
where the dissociation $\ml{C}_{nl}^{(-)}({\bs x}, {\bs k}, t)$ and recombination $\ml{C}_{nl}^{(+)}({\bs x}, {\bs k}, t)$ terms are defined in Eqs.~(\ref{eqn:disso_f}) and (\ref{eqn:reco_f}). Both terms $\ml{C}_{nl}^{(\pm)}({\bs x}, {\bs k}, t)$ consist of three parts: phase space integrals, $\delta$-functions for energy-momentum conservations and scattering amplitudes squared.

The integral over $\diff^3 r$ in $\ml{C}_{nl}^{(+)} $ is nontrivial: not only the wavefunction but also the distribution function is involved. We now consider under what conditions the integral can be further simplified. We note that the support (the region with nonzero function value) of the integrand is on the order of the Bohr radius $a_B$ of the bound state. So if the distribution function is almost uniform in $\bs r$ for $r \lesssim a_B$, one can take the distribution function out of the integral. This is true when the diffusion length scale $\sqrt{Dt}$ is much larger than $r\sim a_B\sim \frac{1}{Mv}$ where $D$ is the diffusion constant of open heavy flavors. The distribution function in ${\bs r}$ caused solely by diffusion is a Gaussian with a variance $\sim Dt$. In other words, the distribution function varies significantly at a length scale $\sqrt{Dt}$ and when one focuses on a region with a much smaller length scale, one can treat the distribution function as uniform. 
A perturbative estimate gives $D\sim\frac{1}{\alpha_s^2T}$ \cite{Moore:2004tg}. The time period for the $Q\bar{Q}$ to be close within the bound state formation range is roughly $t\sim \frac{a_B}{v_\ma{rel}} \sim \frac{1/Mv}{p_\ma{rel}/M}\sim \frac{1}{p_\ma{rel}v}$. So $\sqrt{Dt} \gg \frac{1}{Mv}$ gives
\be
\frac{1}{\alpha_s^2  p_\ma{rel} T}\gg \frac{1}{M^2v}\,.
\ee
Taking $T\sim Mv^2$ as previously assumed we find we must have $p_\ma{rel} \ll M v/(v^2 \alpha_s^2)$ which is clearly satisfied for $p_\ma{rel}\sim Mv$. For  $p_\ma{rel}$ large enough that this condition is not satisfied, the contribution to the integral involving $\langle \psi_{nl}| {\bs r}|\Psi_{{\bs p}_{\ma{rel}}}\rangle$ is negligible for such large $p_\ma{rel}$. This agrees with the intuition: a heavy quark antiquark pair with large relative momentum cannot form a bound state. So we can take $ f_{Q\bar{Q}}(\bs x, {\bs p}_{\ma{cm}}, {\bs r}, {\bs p}_{\ma{rel}}, a,t)$ out of the wavefunction integral with the awareness that the contribution from $r\gg a_B$ should vanish.

Furthermore, we make the molecular chaos assumption and write
\be
 f_{Q\bar{Q}}(\bs x, {\bs p}_{\ma{cm}}, {\bs r}, {\bs p}_{\ma{rel}}, a,t) = \frac{1}{9}f_Q({\bs x}_1, {\bs p}_1, t) f_{\bar{Q}}({\bs x}_2, {\bs p}_2, t)\,,
\ee
where ${\bs x}, {\bs r}, {\bs p}_{\ma{cm}}, {\bs p}_{\ma{rel}}$ are the c.m.~and relative positions and momenta of the heavy quark antiquark pair with positions ${\bs x}_1, {\bs x}_2$ and momenta ${\bs p}_1$, ${\bs p}_2$. The factor $\frac{1}{9}$ accounts for the probability of the color state of $Q\bar{Q}$ being in a specific octet state $a$. The molecular chaos assumption is valid when the rate of decorrelation between the heavy quark and antiquark is much larger than the relaxation rate of the system. The former is given by $D^{-1}\sim \alpha_s^2T$ with $\alpha_s$ at the scale $T$ or $m_D$ while the later has been estimated above and is $\sim \alpha_s v^2T$ with $\alpha_s$ at the scale $Mv$. In NRQCD, $v\sim \alpha_s(Mv)$, so the molecular chaos assumption is valid.

Combining these two assumptions gives
\be\nn
C_{nl}^{(+)}({\bs x}, {\bs k}, t)&=&\frac{8}{9}\int\frac{\diff^3 p_{\ma{cm}}}{(2\pi)^3} \frac{\diff^3 p_{\ma{rel}}}{(2\pi)^3} \frac{\diff^3q}{(2\pi)^32q} (1+n_B(q)) f_Q({\bs x}_1, {\bs p}_1, t) f_{\bar{Q}}({\bs x}_2, {\bs p}_2, t) \\ 
&& (2\pi)^4 \delta^3({\bs k}-{\bs p}_{\ma{cm}}+\bs q) \delta(-|E_{nl}|+q-\frac{{\bs p}_{\ma{rel}}^2}{M}) \frac{2}{3} \frac{T_F}{N_c} q^2g^2 |\langle  \psi_{nl}  | {\bs r} |  \Psi_{{\bs p}_{\ma{rel}}}  \rangle|^2  \,,
\ee
where the sum over color index $a$ has been carried out. This is the recombination term used in the Boltzmann equation in Ref.~\cite{Yao:2017fuc}. In order to take the spin multiplicity into account, one must further insert a factor $g_s$ into $C_{nl}^{(+)}$ where $g_s=\frac{3}{4}$ for $S=1$ quarkonium and $g_s=\frac{1}{4}$ for $S=0$ quarkonium. The in-medium dynamical evolution of open heavy quarks can also be described by Boltzmann equations \cite{Svetitsky:1987gq,Gossiaux:2008jv,Ke:2018tsh}.

\section{Annihilation of quarkonium}
\label{sect:annihilation}
It is known that the NRQCD Lagrangian has four-fermion interactions, which can describe the annihilation of quarkonium (into other hadrons or leptons) and are not included in the pNRQCD Lagrangian Eq.~(\ref{eq:lagr}). 
We can add $-\Gamma S^\dagger S \rho_S $ to describe the annihilation. But this would break the conservation of probability $\Tr(\frac{\diff}{\diff t}\rho_S)=0$. So one also needs to include terms of the forms $\rho_S S^\dagger S$ and $S \rho_S S^\dagger $. A pedagogical  discussion of how to construct an open effective field theory in order to  conserve probability can be found in Ref.~\cite{Braaten:2016sja}. The annihilation is too slow to be of much interest for phenomenology but we study it as an interesting example of how Lindblad-type operators enter the time evolution equation for the density matrix. In our case, we first restore the standard pNRQCD notation of singlet field, $S({\bs R}, {\bs r}, t) \equiv \langle {\bs r} | S({\bs R}, t) \rangle$, i.e., we project the wavefunction of the relative motion onto the relative position space. Then we can add two new terms in the density matrix evolution equation
\be \nn
\rho_S(t) &=& \cdots + \int_0^t dt_1 \int \diff^3R \int \diff^3r \Big( -\frac{\Gamma({\bs r})}{2} \{ S^\dagger({\bs R}, {\bs r}, t_1)S({\bs R}, {\bs r}, t_1), \rho_S(0)  \}
+\\
&& \Gamma({\bs r}) S({\bs R}, {\bs r}, t_1) \rho_S(0) S^\dagger({\bs R}, {\bs r}, t_1)
\Big)\,,
\ee
where the evolution term is explicitly trace-preserving. 

As above, we are interested in the bound state and will sandwich the density matrix between two bound quarkonium states and then do a Wigner transform.

\subsection{$\{ S^\dagger({\bs R}, {\bs r}, t)S({\bs R}, {\bs r}, t), \rho_S(0)  \}$ term} 
We first compute the $S^\dagger({\bs R}, {\bs r}, t)S({\bs R}, {\bs r}, t) \rho_S(0)$ term sandwiched between $ \langle   {\bs k}_1 , nl, 1 |$ and $|   {\bs k}_2 , nl, 1\rangle$ and insert a complete set of states $|  {\bs k}_3, n_3l_3,1 \rangle$.
\be \nn
%&&\int_0^t dt_1 \int \diff^3R \int \diff^3r \frac{-\Gamma({\bs r})}{2}  \langle  {\bs k}_1, nl,1   | S^\dagger_{\bs r}({\bs R}, t_1)S_{\bs r}({\bs R}, t_1) \rho_S(0)  |   {\bs k}_2 , nl, 1\rangle \\ \nn
&&  \int_0^t dt_1 \int \diff^3R \int \diff^3r \frac{-\Gamma({\bs r})}{2} \int\frac{\diff^3k_3}{(2\pi)^3}\sum_{n_3,l_3}   \langle  {\bs k}_1, nl,1   | S^\dagger({\bs R}, {\bs r}, t_1)S({\bs R}, {\bs r}, t_1) |  {\bs k}_3, n_3l_3,1 \rangle \\  \nn
&&\langle {\bs k}_3, n_3l_3,1 | \rho_S(0)  |   {\bs k}_2 , nl, 1\rangle \\ \nn
&=& \int_0^t dt_1 \int \diff^3R \int \diff^3r \frac{-\Gamma({\bs r})}{2} \int\frac{\diff^3k_3}{(2\pi)^3}\sum_{n_3,l_3} \psi_{n_3l_3}({\bs r}) \psi^*_{nl}({\bs r}) 
e^{i|E_{nl}|t_1-i{\bs k}_1\cdot {\bs R}} e^{-i|E_{n_3l_3}|t_1+i{\bs k}_3\cdot {\bs R}} \\ \nn
&&\langle {\bs k}_3, n_3l_3,1 | \rho_S(0)  |   {\bs k}_2 , nl, 1\rangle \\
%&=& \frac{-1}{2} \int\frac{\diff^3k_3}{(2\pi)^3}\sum_{n_3,l_3} \int \diff^3r \,  \psi^*_{nl}({\bs r}) \Gamma({\bs r}) \psi_{n_3l_3}({\bs r}) (2\pi)\delta(E_{k_1}-E_{k_3}) (2\pi)^3\delta^3({\bs k}_1 - {\bs k}_3) \\
%&&\langle {\bs k}_3, n_3l_3,1 | \rho_S(0)  |   {\bs k}_2 , nl, 1\rangle \\ \nn
&=& \frac{-1}{2}t  \int \diff^3r \,  \psi^*_{nl}({\bs r}) \Gamma({\bs r}) \psi_{nl}({\bs r}) \langle {\bs k}_1, nl,1 | \rho_S(0)  |   {\bs k}_2 , nl, 1\rangle \,.
%&\equiv & \frac{-1}{2}t \Gamma_{nl}  \langle {\bs k}_1, nl,1 | \rho_S(0)  |   {\bs k}_2 , nl, 1\rangle \,.
\ee
where we have used the Markovian approximation and written the delta function in energy as $t$. In the summation over $n_3$ and $l_3$, only $n_3=n$, $l_3=l$ contributes due to the delta function in energy (we assume no degeneracy in the bound state eigenenergy beyond that implied by rotational invariance). 

We can define the annihilation rate of a quarkonium state $nl$,
\be
\Gamma_{nl}   \equiv \int \diff^3r \,  \psi^*_{nl}({\bs r}) \Gamma({\bs r}) \psi_{nl}({\bs r})\,,
\ee
which is nonzero even in vaccum and should be distinguished from the dissociation rate inside QGP.
For $S$-wave, one may set $\Gamma({\bs r}) = \Gamma \delta^3(\bs r)$ and then $\Gamma_S =  \Gamma |\psi_{S}(0)|^2$, i.e., the annihilation rate depends on the wavefunction of the relative motion at the origin.

The other term in the anticommutator will give the same result. Under a Wigner transform, these two terms lead in the Boltzmann equations to
\be
\label{eqn:annhi1}
- \Gamma_{nl}  f_{nl}({\bs x}, {\bs k}, t)\,.
\ee
Typically $\Gamma_{nl} \sim 10 $ keV, so for a QGP with a lifetime $\sim10$ fm $\sim0.05$ MeV$^{-1}$ , the effect from quarkonium annihilations is negligible on the in-medium evolution. It is justified to assume that the total number of heavy quarks is conserved during the in-medium evolution.

\subsection{$S({\bs R}, {\bs r}, t) \rho_S(0) S^\dagger({\bs R}, {\bs r}, t)$ term}
Then we compute the contribution from the $S({\bs R}, {\bs r}, t) \rho_S(0) S^\dagger({\bs R}, {\bs r}, t)$ term sandwiched between $ \langle   {\bs k}_1 , nl, 1 |$ and $|   {\bs k}_2 , nl, 1\rangle$:
\be  \nn
&&\int_0^t dt_1 \int \diff^3R \int \diff^3r \Gamma({\bs r}) \langle  {\bs k}_1, nl, 1   |   S({\bs R}, {\bs r}, t_1) \rho_S(0)  S^\dagger({\bs R}, {\bs r}, t_1)   |   {\bs k}_2 , nl, 1\rangle \\ \nn
%&=& \int_0^t dt_1 \int \diff^3R \int \diff^3r \Gamma({\bs r}) \int\frac{\diff^3k_3}{(2\pi)^3}\sum_{n_3,l_3} \int\frac{\diff^3k_4}{(2\pi)^3}\sum_{n_4,l_4} \psi_{n_3l_3}({\bs r})  \psi^*_{n_4l_4}({\bs r}) \\
%&&e^{iE_{k_4}t_1-i{\bs k}_4\cdot {\bs R}} e^{-iE_{k_3}t_1+i{\bs k}_3\cdot {\bs R}} 
%\langle    {\bs k}_1, nl, 1;\ {\bs k}_3, n_3l_3, 1    | \rho_S(0) |       {\bs k}_2, nl, 1;\ {\bs k}_4, n_4l_4, 1     \rangle \\\nn
&=& \int \diff^3r \Gamma({\bs r}) \int\frac{\diff^3k_3}{(2\pi)^3}\sum_{n_3,l_3} \int\frac{\diff^3k_4}{(2\pi)^3}\sum_{n_4,l_4} \psi_{n_3l_3}({\bs r})  \psi^*_{n_4l_4}({\bs r})
 (2\pi)^4\delta^3({\bs k}_3 - {\bs k}_4) \delta(E_{k_3}-E_{k_4}) \\ \nn
&&\langle    {\bs k}_1, nl, 1;\ {\bs k}_3, n_3l_3, 1    | \rho_S(0) |       {\bs k}_2, nl, 1;\ {\bs k}_4, n_4l_4, 1     \rangle \\\nn
%&=& t \int\frac{\diff^3k_3}{(2\pi)^3}\sum_{n_3,l_3}  \int \diff^3r \ \psi^*_{n_3l_3}({\bs r}) \Gamma({\bs r}) \psi_{n_3l_3}({\bs r}) \langle    {\bs k}_1, nl, 1;\ {\bs k}_3, n_3l_3, 1    | \rho_S(0) |       {\bs k}_2, nl, 1;\ {\bs k}_3, n_3l_3, 1     \rangle \\ \nn
%&=& t \int\frac{\diff^3k_3}{(2\pi)^3}\sum_{n_3,l_3}  \Gamma_{n_3l_3}   \int\frac{\diff^3k'_3}{(2\pi)^3} \delta^3({\bs k}_3') \\
%&& \langle    {\bs k}_1, nl, 1;\ {\bs k}_3 + \frac{{\bs k}_3'}{2}, n_3l_3, 1    | \rho_S(0) |       {\bs k}_2, nl, 1;\ {\bs k}_3-\frac{{\bs k}_3'}{2}, n_3l_3, 1     \rangle \\ \nn
&=& t \int\frac{\diff^3k_3}{(2\pi)^3}\sum_{n_3,l_3}  \Gamma_{n_3l_3}   \int\frac{\diff^3k'_3}{(2\pi)^3} \int \diff^3x' e^{i{\bs k}_3'\cdot {\bs x}'} \\
&&\langle    {\bs k}_1, nl, 1;\ {\bs k}_3 + \frac{{\bs k}_3'}{2}, n_3l_3, 1    | \rho_S(0) |       {\bs k}_2, nl, 1;\ {\bs k}_3-\frac{{\bs k}_3'}{2}, n_3l_3, 1  \rangle \,,
\ee
where we have inserted an identity $\int\diff^3k'_3\delta^3({\bs k}'_3)=1$ and written $\delta^3({\bs k}'_3)$ as a spatial integral over ${\bs x}'$. It should be noted that the integral over ${\bs k}_3'$ is already a Wigner transform on the density matrix of the second particle with momentum ${\bs k}_3$ and position ${\bs x}'$. If we further apply a Wigner transform on the density matrix of the first particle and properly reshuffle labels, we obtain the contribution of this term in the Boltzmann equation
\be
 \sum_{n',l' } \Gamma_{n'l'}   \int\frac{\diff^3k'}{(2\pi)^3} \int \diff^3x'  f({\bs x}, {\bs k}, nl;\ {\bs x}', {\bs k}', n'l';\ t)\,.
\ee
It involves the two-particle distribution function $ f({\bs x}, {\bs k}, nl;\ {\bs x}', {\bs k}', n'l';\ t)$ of two quarkonium states $nl$ and $n'l'$ with positions ${\bs x}$, ${\bs x}'$ and momenta ${\bs k}$, ${\bs k}'$ respectively. When the second quarkonium with the quantum number $n'l'$ annihilates, it leads to an increase in the one-particle distribution function of quarkonium with quantum number $nl$. Therefore, this term together with the term in Eq.~(\ref{eqn:annhi1}) guarantees the conservation of probability in the one-particle distribution of quarkonium. However, as mentioned earlier,  the annihilation effect is negligible in current heavy ion collision experiments.

\section{Conclusion}
\label{sect:conclusion}
In this paper, we used the open quantum system formalism where the system of heavy quarks and quarkonium is described by pNRQCD at LO in the non-relativistic expansion while the environment is a weakly-coupled thermal QGP. We derived the Boltzmann transport equation for a quarkonium inside a QGP below its melting temperature from first principles under the assumptions of weak coupling between the quarkonium and the QGP and Markovian evolution.  Both assumptions are justified by our assumed separation of scales, $M\gg Mv \gg Mv^2 \gtrsim T \gtrsim m_D$. This requires $rT\ll 1$, where $r$ is the typical size of the quarkonium, which is probably a realistic assumption for current heavy ion collisions for the most compact quarkonium such as the $\Upsilon$(1S). It could work as well for $\Upsilon$(2S) for $T \lesssim 200$ MeV.
Correlations of environment operators are calculated in real-time thermal field theory. After tracing out the environment degrees of freedom, we obtained the Lindblad equation, which is non-unitary and time-irreversible. Under a Wigner transform, the Lindblad equation leads to the Boltzmann transport equation. 

The derivation here provides a theoretical justification of quarkonium transport equations inside a weakly-coupled QGP below the melting temperature of the quarkonium. It connects two main approaches of the phenomenology of quarkonium production in heavy ion collisions. One can improve the derivation by working to next-leading-order in the coupling constant and expansion parameters in both the system sector (pNRQCD) and the environment sector (thermal QCD). In the case of a non-perturbative construction of pNRQCD, a similar derivation is possible. The connection between the complex potential calculated on the lattice \cite{Rothkopf:2011db} and the transport equation is worth exploring in our framework. The derivation can be extended to the case of quarkonium evolution inside a strongly-coupled QGP, a hot medium out of equilibrium or cold nuclear matter by replacing the Green's functions of thermal QCD with those in the corresponding media. It would also be interesting to study the viscous and anisotropic corrections to the Debye screening, the dissociation rate of quarkonium \cite{Dumitru:2007hy,Dumitru:2009fy,Du:2016wdx} and the recombination in a non-thermal QGP. The effect of a turbulent plasma on the heavy quark antiquark pair or quarkonium in the early stage of heavy ion collisions can also be explored \cite{Mrowczynski:2017kso}. A description of the quarkonium evolution through cold nuclear matter will be useful to studies of quarkonium production in both proton-ion and electron-ion collisions.

\begin{acknowledgements}
We thank Berndt M\"uller for helpful discussions. X.Y. is supported by U.S. Department of Energy research grant DE-FG02-05ER41367 and Brookhaven National Laboratory. T.M. is supported by U.S. Department of Energy research grant DE-FG02-05ER41368.
\end{acknowledgements}

\appendix
\section{Green's functions in real-time thermal field theory}
\label{app:real-time}

In real-time thermal field theory, the commonly used Green's functions are the ``$>$", ``$<$", retarded, advanced and time-ordered Green's functions. They are defined as follows:
\be \nn
D_{\mu\nu}^{>\,ab}(t,\bs x)&=& \big\langle A^a_{\mu}(t,\bs x)A^b_{\nu}(0,0)   \big\rangle _T \\ \nn
D_{\mu\nu}^{<\,ab}(t,\bs x) &=& \big\langle  A^a_{\nu}(0,0) A^b_{\mu}(t,\bs x) \big\rangle_T\\ \nn
D_{\mu\nu}^{R\,ab} (t,\bs x)&=& \big\langle  \theta(t)[A^a_{\mu}(t,\bs x), A^b_{\nu}(0,0)  ] \big\rangle_T\\ \nn
D_{\mu\nu}^{A\,ab} (t,\bs x)&=& - \big\langle  \theta(-t) [A^a_{\mu}(t,\bs x), A^b_{\nu}(0,0)  ] \big\rangle_T\\
D^{T\,ab}_{\mu\nu} (t,\bs x) &=& \big\langle  \ml{T}(A^a_{\mu}(t,\bs x)A^b_{\nu}(0,0)  ) \big\rangle_T\,.
\ee
The Fourier transform is defined as
\be
D_{\mu\nu}^{X\,ab} (q_0,\bs q) = \int \diff^4x e^{i(q_0t-\bs q\cdot \bs x)}D^{X\,ab}_{\mu\nu}(t,\bs x)\,,
\ee
where $X$ could be $>$, $<$, $R$, $A$ or $\ml{T}$. In momentum space for a free theory
\be \nn
\label{eqn:>}
D_{\mu\nu}^{>\,ab}(q) &=& (1+n_B(q_0))\sum_{\lambda}\epsilon_{\mu}^{\lambda*}\epsilon_{\nu}^{\lambda} \delta^{ab} \rho_F(q)\\ \nn
\label{eqn:<}
D_{\mu\nu}^{<\,ab}(q) &=& n_B(q_0)\sum_{\lambda}\epsilon_{\mu}^{\lambda*}\epsilon_{\nu}^{\lambda} \delta^{ab} \rho_F(q)\\ \nn
\rho_F(q) &=&  (2\pi)\sign(q_0) \delta(q_0^2-{\bs q}^2)\\ \nn
D_{\mu\nu}^{R\,ab} (q)&=& \frac{i\sum_{\lambda}\epsilon_{\mu}^{\lambda*}\epsilon_{\nu}^{\lambda} \delta^{ab} }{q_0^2-{\bs q}^2+i\sign(q_0)\epsilon} \\ \nn
D_{\mu\nu}^{A\,ab} (q)&=& \frac{i\sum_{\lambda}\epsilon_{\mu}^{\lambda*}\epsilon_{\nu}^{\lambda} \delta^{ab} }{q_0^2-{\bs q}^2-i\sign(q_0)\epsilon} \\ \nn
D^{T\,ab}_{\mu\nu} (q) &=& D_{\mu\nu}^{Rab} (q) + D_{\mu\nu}^{<ab}(q)\\
&=& \sum_{\lambda}\epsilon_{\mu}^{\lambda*}\epsilon_{\nu}^{\lambda} \delta^{ab}\bigg( \frac{i }{q_0^2-{\bs q}^2+i\epsilon} + n_B(|q_0|)(2\pi)\delta(q_0^2-{\bs q}^2) \bigg)\,.
\ee

\section{$-i\sigma_{ab} [L_{ab},\rho_S]$ term}
\label{app:static}
To compute $\sum_{a,b}\sigma_{ab} L_{ab}$, we first note that 
\be \nn
\label{eqn:sigma_ab_theta}
 && \bigg( \sum_{a,b}\frac{-i}{2}\int_0^t \diff t_1 \int_0^t  \diff t_2 \sum_{\alpha,\beta} C_{\alpha\beta}(t_1,t_2) \theta(t_1-t_2) \langle a | O^{(S)}_{\alpha}(t_1) O^{(S)}_\beta(t_2) | b \rangle L_{ab}\bigg)^\dagger\\ \nn
 &=& \sum_{a,b}\frac{i}{2}\int_0^t \diff t_1 \int_0^t  \diff t_2 \sum_{\alpha,\beta} C_{\beta\alpha}(t_2,t_1) \theta(t_1-t_2) \langle b |  O^{(S)}_\beta(t_2) O^{(S)}_{\alpha}(t_1) | a \rangle L_{ba}\\ 
  &=& \sum_{a,b}\frac{-i}{2}\int_0^t \diff t_1 \int_0^t  \diff t_2 \sum_{\alpha,\beta} C_{\alpha\beta}(t_1,t_2) \big(-\theta(t_2-t_1)\big) \langle a | O^{(S)}_{\alpha}(t_1) O^{(S)}_\beta(t_2) | b \rangle L_{ab}\,,
\ee
where in the last line we flipped $\alpha\leftrightarrow\beta$, $\ t_1\leftrightarrow t_2$ and $\ |a\rangle \leftrightarrow |b\rangle$. We can split $\sign(t_1-t_2)$ into $\theta(t_1-t_2)$ and $-\theta(t_2-t_1)$ in $\sum_{a,b}\sigma_{ab} L_{ab}$ and just need to compute the $\theta(t_1-t_2)$ term. The $-\theta(t_2-t_1)$ term is given by the Hermitian conjugate of the $\theta(t_1-t_2)$ term. Therefore $\sum_{a,b}\sigma_{ab}L_{ab}$ is Hermitian and this term can be thought of as a correction to the system Hamiltonian.  To carry out the calculation explicitly, we first write
\be 
D^{>\,ab}_{\mu\nu}({\bs R}_1,t_1; {\bs R}_2, t_2)\theta(t_1-t_2) = \big( D^{R\,ab}_{\mu\nu}({\bs R}_1,t_1; {\bs R}_2, t_2) + D^{<\,ab}_{\mu\nu}({\bs R}_1,t_1; {\bs R}_2, t_2)  \big) \theta(t_1-t_2)\,. \,\,\,\,
\ee
Then we can replace $C_{\alpha\beta}(t_1,t_2)$ in the first line of Eq.~(\ref{eqn:sigma_ab_theta}), due to the $\theta(t_1-t_2)$, with 
\be \nn
C_{\alpha\beta}(t_1,t_2) &=& 
C_{{\bs R}_1i_1a_1,{\bs R}_2i_2a_2}(t_1, t_2) = \frac{T_F}{N_C}g^2\langle  E_{i_1}^{a_1}({\bs R}_1, t_1)  E_{i_2}^{a_2}({\bs R}_2, t_2)  \rangle_T \\ \nn
&\rightarrow &
%\frac{T_F}{N_C}g^2 \delta^{a_1a_2} \int\frac{\diff^4q}{(2\pi)^4}  e^{-iq_0(t_1-t_2) + i\bs q\cdot({\bs R}_1-{\bs R}_2)}  \\ \nn
%&& (q_0^2\delta_{i_1i_2} -q_{i_1}q_{i_2}) \bigg[ \frac{i}{q_0^2 -{\bs q}^2 + i\sign(q_0)\epsilon} +   n_B(q_0)(2\pi) \sign(q_0) \delta(q_0^2-{\bs q}^2)  \bigg]\\
\frac{T_F}{N_C}g^2 \delta^{a_1a_2} \int\frac{\diff^4q}{(2\pi)^4}  e^{-iq_0(t_1-t_2) + i\bs q\cdot({\bs R}_1-{\bs R}_2)}  \\ 
\label{eqn:real1}
&& (q_0^2\delta_{i_1i_2} -q_{i_1}q_{i_2})  \bigg[ \frac{i}{q_0^2 -{\bs q}^2 + i\epsilon} +   n_B(|q_0|)(2\pi)  \delta(q_0^2-{\bs q}^2)  \bigg] + \ml{O}(g^3)\,.
\ee
The term inside the square brackets is the time-ordered thermal propagator in momentum space.

We are interested in the bound state part of the density matrix, so we set $|a\rangle = |{\bs k}_1, n_1l_1, 1\rangle$ and $|b\rangle = | {\bs k}_2, n_2l_2, 1 \rangle$. Then we can compute
\be  \nn
&& \theta(t_1-t_2) \langle a | O^{(S)}_{\alpha}(t_1) O^{(S)}_\beta(t_2) | b \rangle \\ \nn
&=&  \theta(t_1-t_2) \langle {\bs k}_1, n_1l_1, 1|   \langle S({\bs R}_1, t_1) | r_{i_1} | O^{a_1}({\bs R}_1, t_1) \rangle 
 \langle O^{a_2}({\bs R}_2, t_2) | r_{i_2} | S({\bs R}_2, t_2) \rangle |  {\bs k}_2, n_2l_2, 1 \rangle \\ \nn
&=&  \delta^{a_1a_2}  \int \frac{\diff^3p_\ma{cm} }{(2\pi)^3}\int \frac{\diff^3p_\ma{rel} }{(2\pi)^3} \langle \psi_{n_1l_1} | r_{i_1} | \Psi_{{\bs p}_\ma{rel}} \rangle
\langle \Psi_{{\bs p}_\ma{rel}} | r_{i_2} |  \psi_{n_2l_2} \rangle \\ 
&&  \theta(t_1-t_2) e^{iE_{p}(t_2-t_1) - i {\bs p}_\ma{cm}\cdot ({\bs R}_2- {\bs R}_1)}  e^{-iE_{k_2}t_2 + i{\bs k}_2\cdot {\bs R}_2} e^{iE_{k_1}t_1 - i{\bs k}_1\cdot {\bs R}_1}  \,,
\ee
where $E_p = \frac{{\bs p}_{\ma{rel}}^2}{M}$. This can be written as 
\be \nn
&& \theta(t_1-t_2) \langle a | O^{(S)}_{\alpha}(t_1) O^{(S)}_\beta(t_2) | b \rangle \\ \nn
 &=&   \langle \psi_{n_1l_1} | r_{i_1}    \int \frac{\diff^4p_\ma{cm} }{(2\pi)^4}\int \frac{\diff^3p_\ma{rel} }{(2\pi)^3}  \frac{i  | \Psi_{{\bs p}_\ma{rel}} \rangle  \langle \Psi_{{\bs p}_\ma{rel}} |}{p_{\ma{cm}}^0-E_{p}+i\epsilon}   r_{i_2} |  \psi_{n_2l_2} \rangle \\ 
\label{eqn:real2}
&&  \delta^{a_1a_2} e^{ip_\ma{cm}^0(t_2-t_1) - i {\bs p}_\ma{cm}\cdot ({\bs R}_2- {\bs R}_1)}  e^{-iE_{k_2}t_2 + i{\bs k}_2\cdot {\bs R}_2} e^{iE_{k_1}t_1 - i{\bs k}_1\cdot {\bs R}_1}\,.
\ee
It should be noted that $p_{\ma{cm}}^0$ here does not represent the c.m.~energy of the octet. In fact, it is the total energy of the composite octet particle, $p_{\ma{cm}}^0 = \frac{{\bs p}_{\ma{cm}}^2}{4M} + \frac{{\bs p}_{\ma{rel}}^2}{M} = \frac{{\bs p}_{\ma{rel}}^2}{M} + \ml{O}(Mv^4)$.

To simplify the expression, we make the Markovian approximation $t\rightarrow\infty$. Then integrating over $t_1$ and $t_2$ will give two $\delta$-functions in energy. Plugging Eqs.~(\ref{eqn:real1}) and (\ref{eqn:real2}) into $\sum_{a,b}\sigma_{ab}L_{ab}$ and integrating over $t_1$, $t_2$, ${\bs R}_1$ and ${\bs R}_2$ we find 
\be  \nn
&&\sum_{a,b}\sigma_{ab} L_{ab} =\frac{1}{2}    \bigg\{ -i \sum_{n_1,l_1}\sum_{n_2,l_2} \sum_{i_1,i_2}\int\frac{\diff^3k_1}{(2\pi)^3}\int\frac{\diff^3k_2}{(2\pi)^3}\int\frac{\diff^4q}{(2\pi)^4} 
\int\frac{\diff^4 p_\ma{cm}}{(2\pi)^4} \int\frac{\diff^3 p_\ma{rel}}{(2\pi)^3} \\ \nn
&&  \frac{T_F}{N_C}(N_C^2-1) g^2  (q_0^2 \delta_{i_1i_2} -q_{i_1}q_{i_2}) \Big( \frac{i}{q_0^2 - {\bs q}^2 + i\epsilon} +   n_B(|q_0|)(2\pi)  \delta(q_0^2-{\bs q}^2)  \Big) \\ \nn
&&  \langle \psi_{n_1l_1} | r_{i_1}  \frac{i  | \Psi_{{\bs p}_\ma{rel}} \rangle  \langle \Psi_{{\bs p}_\ma{rel}} |}{p_{\ma{cm}}^0-E_{p}+i\epsilon}   r_{i_2} |  \psi_{n_2l_2} \rangle \\ \nn
&& (2\pi)^3\delta^3 ({\bs k}_1-{\bs p}_\ma{cm} - {\bs q}) (2\pi)^3\delta^3 ({\bs k}_2-{\bs p}_\ma{cm} - {\bs q}) \\\nn
&& (2\pi)\delta(E_{k_1}-p_\ma{cm}^0-q^0) (2\pi)\delta(E_{k_2}-p_\ma{cm}^0-q^0)\\
&& L_{| {\bs k}_1,n_1l_1,1 \rangle \langle  {\bs k}_2,n_2l_2,1   |} +\ma{h.c.}\bigg\}\,.
\ee
The two time integrals give a product of two delta functions in energy $\delta(\omega_1)\delta(\omega_2)$, where $\omega_i=E_{k_i}-p_\ma{cm}^0-q^0$ for $i=1,2$ and $\omega_1=\omega_2 =\omega$. We interpret one factor of $2\pi\delta(\omega)$ to be the time interval, so the double time integral is interpreted as follows
\be
\int_0^t\diff t_1\int_0^t \diff t_2 e^{i\omega t_1}e^{-i\omega t_2} = \frac{4\sin^2(\omega t/2)}{\omega^2} \xrightarrow{t\to \infty} t2\pi\delta(\omega)\,.
\ee
See Ref.~\cite{JJSakurai} for details.
This argument also applies in Appendixes \ref{app:dissociation} and \ref{app:recombination}. The $\delta$-functions in energy and momentum give ${\bs k}_1={\bs k}_2={\bs k}$, $n_1=n_2=n$ and $l_1=l_2=l$ (we assume no degeneracy in the bound state eigenenergy beyond that implied by rotational invariance). So we have
\be \nn
\sum_{a,b}\sigma_{ab}L_{ab} &\rightarrow& t \sum_{n,l} \int\frac{\diff^3k}{(2\pi)^3} \Re  \bigg\{ -i g^2 C_F  \sum_{i_1,i_2} \int\frac{\diff^4q}{(2\pi)^4} 
\int\frac{\diff^4 p_\ma{cm}}{(2\pi)^4} \int\frac{\diff^3 p_\ma{rel}}{(2\pi)^3} \\ \nn
&& (q_0^2 \delta_{i_1i_2} -q_{i_1}q_{i_2}) \Big( \frac{i}{q_0^2 - {\bs q}^2 + i\epsilon} +   n_B(|q_0|)(2\pi)  \delta(q_0^2-{\bs q}^2)  \Big) \\ \nn
&& \langle \psi_{nl} | r_{i_1}  \frac{i  | \Psi_{{\bs p}_\ma{rel}} \rangle  \langle \Psi_{{\bs p}_\ma{rel}} |}{p_{\ma{cm}}^0-E_{p}+i\epsilon}   r_{i_2} |  \psi_{nl} \rangle\\ 
&& (2\pi)^3\delta^3 ({\bs k}-{\bs p}_\ma{cm} - {\bs q}) (2\pi)\delta(E_{k}-p_\ma{cm}^0-q^0) 
\bigg\} L_{| {\bs k},nl,1 \rangle \langle  {\bs k},nl,1   |} \,,
\ee
where the terms for the unbound singlet and octet are not shown here but can be written out similarly.

\section{$-\frac{1}{2}\gamma_{ab,cd}\{L^{\dagger}_{cd}L_{ab}, \rho_S(0)\}$ term}
\label{app:dissociation}
We will show this term gives the dissociation term in the Boltzmann equation. We first compute the term $-\frac{1}{2}\gamma_{ab,cd}L^{\dagger}_{cd}L_{ab} \rho_S(0)$\,. As explained previously, we are interested in the bound state part of the density matrix and take $\langle {\bs k}_1, n_1l_1, 1 | \rho_S(t) | {\bs k}_2, n_2l_2, 1\rangle $, so we set $|d\rangle=|{\bs k}_1, n_1l_1, 1 \rangle$. Since at lowest order of the expansion, the transition between the bound state and unbound pair only occurs via singlet-octet transition, we have $|a\rangle=|c\rangle=|{\bs p}_{\ma{cm}}, {\bs p}_{\ma{rel}}, a_1\rangle$ and double summations over $|a\rangle$ and $|c\rangle$ become just one summation. Similarly we have $|b\rangle = |{\bs k}_3, n_3l_3, 1\rangle$. We need to compute
\be
\gamma_{ab,cd} &=& \int \diff^3R_1 \int \diff^3R_2 \sum_{i_1,i_2,b_1,b_2}\int_0^t \diff t_1 \int_0^t \diff t_2 C_{{\bs R}_1i_1b_1,{\bs R}_2i_2b_2}(t_1, t_2) \\ \nn
&&\langle {\bs k}_1, n_1l_1, 1 | \langle S({\bs R}_1, t_1) | r_{i_1} | O^{b_1}({\bs R}_1, t_1) \rangle | {\bs p}_{\ma{cm}}, {\bs p}_{\ma{rel}},a_1\rangle 
%&&\langle {\bs k}_1, n_1l_1, 1 | S^{\dagger}({\bs R}_1, t_1)  r_{i_1}  O^{b_1}({\bs R}_1, t_1)  | {\bs p}_{\ma{cm}}, {\bs p}_{\ma{rel}},a_1\rangle 
\\\nn
&&\langle {\bs p}_{\ma{cm}}, {\bs p}_{\ma{rel}},a_1 |  \langle O^{b_2}({\bs R}_2, t_2) | r_{i_2} | S({\bs R}_2, t_2) \rangle | {\bs k}_3, n_3l_3, 1\rangle \,.
%&&\langle {\bs p}_{\ma{cm}}, {\bs p}_{\ma{rel}},a_1 |   O^{b_2\dagger}({\bs R}_2, t_2)  r_{i_2} S({\bs R}_2, t_2) | {\bs k}_3, n_3l_3, 1\rangle \,.
\ee
We can start with
\be \nn
%&&\langle {\bs k}_1, n_1l_1, 1 | S^{\dagger}({\bs R}_1, t_1) r_{i_1} O^{b_1}({\bs R}_1, t_1)  | {\bs p}_{\ma{cm}}, {\bs p}_{\ma{rel}},a_1\rangle \\\nn
&&\langle {\bs k}_1, n_1l_1, 1 | \langle S({\bs R}_1, t_1) | r_{i_1} | O^{b_1}({\bs R}_1, t_1) \rangle | {\bs p}_{\ma{cm}}, {\bs p}_{\ma{rel}},a_1\rangle \\\nn
  &=&  \langle \psi_{n_1l_1} | r_{i_1}| \Psi_{{\bs p}_{\ma{rel}}} \rangle \delta^{a_1b_1}e^{-i(E_{{\bs p}}t_1 - {\bs p}_{\ma{cm}} \cdot {\bs R}_1)} e^{i(E_{{\bs k}_1}t_1 - {\bs k}_1\cdot{\bs R}_1)} \\
%&&\langle {\bs p}_{\ma{cm}}, {\bs p}_{\ma{rel}},a_1 |  O^{b_2\dagger}({\bs R}_2, t_2) r_{i_2} S({\bs R}_2, t_2) | {\bs k}_3, n_3l_3, 1\rangle \\\nn
&&\langle {\bs p}_{\ma{cm}}, {\bs p}_{\ma{rel}},a_1 |  \langle O^{b_2}({\bs R}_2, t_2) | r_{i_2} | S({\bs R}_2, t_2) \rangle | {\bs k}_3, n_3l_3, 1\rangle \\\nn
 &=&  \langle \Psi_{{\bs p}_{\ma{rel}}}   | r_{i_2}|  \psi_{n_3l_3} \rangle \delta^{a_1b_2}e^{-i(E_{{\bs k}_3}t_2 - {\bs k}_3\cdot{\bs R}_2 )} e^{i(E_{{\bs p}}t_2 - {\bs p}_{\ma{cm}} \cdot {\bs R}_2)} \,,
\ee
where $E_{{\bs p}} = \frac{{\bs p}_{\ma{rel}}^2}{M}$ and $E_{{\bs k}_i} = -|E_{n_il_i}|$ up to $v^2$-corrections. The correlation needed is 
\be\nn
&&C_{{\bs R}_1i_1b_1,{\bs R}_2i_2b_2}(t_1, t_2) = \frac{T_F}{N_C}g^2\langle  E_{i_1}^{b_1}({\bs R}_1, t_1)  E_{i_2}^{b_2}({\bs R}_2, t_2)  \rangle_T \\ \nn
&=&\frac{T_F}{N_C}g^2 \delta^{b_1b_2} \int\frac{\diff^4q}{(2\pi)^4}  e^{iq_0(t_1-t_2)-i\bs q\cdot({\bs R}_1-{\bs R}_2)} \\
&& (q_0^2\delta_{i_1i_2} -q_{i_1}q_{i_2}) n_B(q_0) (2\pi) \sign(q_0) \delta(q_0^2-{\bs q}^2) + \ml{O}(g^3) \,,
\ee
where we have used the expression of $D^{<\,ab}_{\mu\nu}(q)$ in Appendix~\ref{app:real-time}. It should be pointed out that one can also use the expression of $D^{>\,ab}_{\mu\nu}(q)$ and will obtain the same result due to the relation $1+n_B(q_0)+n_B(-q_0) = 0$. Now we can write the term $\langle {\bs k}_1, n_1l_1, 1 | \gamma_{ab,cd}L^{\dagger}_{cd}L_{ab} \rho_S(0)  | {\bs k}_2, n_2l_2, 1\rangle$ out explicitly as
\be\nn
&&\int\frac{\diff^3p_{\ma{cm}}}{(2\pi)^3} \frac{\diff^3p_{\ma{rel}}}{(2\pi)^3} \frac{\diff^3k_{3}}{(2\pi)^3} \frac{\diff^4q}{(2\pi)^4} \int \diff^3R_1 \int \diff^3R_2 \int_0^t\diff t_1 \int_0^t\diff t_2 \sum_{n_3,l_3,a_1,b_1,b_2,i_1,i_2}  \\ \nn
&& \frac{T_F}{N_C}g^2 (q_0^2\delta_{i_1i_2} -q_{i_1}q_{i_2})  n_B(q_0)\delta^{b_1b_2}(2\pi) \sign(q_0) \delta(q_0^2-{\bs q}^2) e^{iq_0(t_1-t_2)-i\bs q\cdot({\bs R}_1-{\bs R}_2)} \\ \nn
&& \langle \psi_{n_1l_1} | r_{i_1}| \Psi_{{\bs p}_{\ma{rel}}} \rangle \delta^{a_1b_1}  e^{-i(E_{{\bs p}}t_1 - {\bs p}_{\ma{cm}} \cdot {\bs R}_1)} e^{i(E_{{\bs k}_1}t_1 - {\bs k}_1\cdot{\bs R}_1)}  \\ \nn
&&  \langle \Psi_{{\bs p}_{\ma{rel}}}   | r_{i_2}|  \psi_{n_3l_3} \rangle \delta^{a_1b_2}  e^{- i(E_{{\bs k}_3}t_2 - {\bs k}_3\cdot{\bs R}_2 )} e^{i(E_{{\bs p}}t_2 - {\bs p}_{\ma{cm}} \cdot {\bs R}_2)}  \\ 
&&  \langle {\bs k}_3, n_3l_3, 1| \rho_S(0) | {\bs k}_2, n_2l_2, 1\rangle \,.
\ee
Integrating over ${\bs R}_1$ and ${\bs R}_2$ gives two delta functions in momenta, $\delta^3({\bs k}_1-{\bs p}_{\ma{cm}}+\bs q)\delta^3({\bs k}_3-{\bs p}_{\ma{cm}}+\bs q)$. Under the Markovian approximation, $t\rightarrow\infty$, integrating over $t_1$ and $t_2$ will give another two delta functions $\delta(E_{k_1}-E_{p}+q_0)\delta(E_{k_3}-E_{p}+q_0)$. Since $E_{k_i}<0$ and $E_{p}>0$, some energy has to be transferred to the bound state to break it up to an unbound state, and thus $q_0$ has to be positive. (If we use $D^{>\,ab}_{\mu\nu}(q)$, here we would have $\delta(E_{k_1}-E_{p}-q_0)\delta(E_{k_3}-E_{p}-q_0)$ and $q_0$ is negative.)

Integrating over ${\bs R}_1$, ${\bs R}_2$, $t_1$, $t_2$ and ${\bs k}_3$ gives
\be \nn
&&\sum_{n_3,l_3,a_1,i_1}\int\frac{\diff^3p_{\ma{cm}}}{(2\pi)^3} \frac{\diff^3p_{\ma{rel}}}{(2\pi)^3} \frac{\diff^3q}{(2\pi)^32q} n_B(q) (2\pi)^5\delta(E_{k_1}-E_{p}+q)\delta(E_{k_3}-E_{p}+q) \delta^3({\bs k}_1 - {\bs p}_{\ma{cm}} +{\bs q} )\\
 &&   \frac{2T_F}{3N_C}q^2g^2  \langle \psi_{n_1l_1} | r_{i_1} | \Psi_{{\bs p}_{\ma{rel}}} \rangle   \langle \Psi_{{\bs p}_{\ma{rel}}}   | r_{i_1} |  \psi_{n_3l_3} \rangle   \langle {\bs k}_1, n_3l_3, 1| \rho_S(0) | {\bs k}_2, n_2l_2, 1\rangle \,,
\ee
where we have used for any smooth function $f(q)$
\be
\int \frac{\diff^3q}{(2\pi)^3} (q^2\delta_{i_1i_2} -q_{i_1}q_{i_2})  f(q) = \frac{2}{3} \delta_{i_1i_2} \int \frac{\diff^3q}{(2\pi)^3} q^2  f(q) \,.
\ee
Due to the energy $\delta$-functions, the sum over $n_3$ and $l_3$ gives $n_3=n_1$, $l_3=l_1$ (we assume no degeneracy in the bound state eigenenergy beyond that implied by rotational invariance). Then one of the energy $\delta$-functions multiplied by $2\pi$ can be interpreted as the time interval $t$. So we have
\be \nn
&&t\int\frac{\diff^3p_{\ma{cm}}}{(2\pi)^3} \frac{\diff^3p_{\ma{rel}}}{(2\pi)^3} \frac{\diff^3q}{(2\pi)^32q} n_B(q)  (2\pi)^4\delta^3({\bs k}_1 - {\bs p}_{\ma{cm}}  +{\bs q} )  \delta(E_{k_1}-E_{p}+q) \\
\label{eqn:disso_rho}
 && \frac{2T_F}{3N_C}(N_C^2-1)q^2g^2 | \langle \psi_{n_1l_1} | \bs r | \Psi_{{\bs p}_{\ma{rel}}} \rangle|^2  \langle {\bs k}_1, n_1l_1, 1| \rho_S(0) | {\bs k}_2, n_2l_2, 1\rangle \,.
\ee
Under a Wigner transform of the form Eq.~(\ref{eqn:wigner}) (where we set ${\bs k}_1 = {\bs k} + \frac{{\bs k}'}{2}$, ${\bs k}_2 = {\bs k}-\frac{{\bs k}'}{2}$,  $n_1=n_2=n$ and $l_1=l_2=l$ and then do a shift in c.m.~momentum ${\bs p}_{\ma{cm}} \rightarrow {\bs p}_{\ma{cm}} + \frac{{\bs k}'}{2}$), Eq.~(\ref{eqn:disso_rho}) finally leads to
\be \nn
 &&t\int\frac{\diff^3p_{\ma{cm}}}{(2\pi)^3} \frac{\diff^3p_{\ma{rel}}}{(2\pi)^3} \frac{\diff^3q}{(2\pi)^32q} n_B(q)  (2\pi)^4\delta^3({\bs k} - {\bs p}_{\ma{cm}} +{\bs q} ) \delta(E_{k}-E_{p}+q)  \\
 \label{eqn:disso_boltz}
 && \frac{2}{3}C_Fq^2g^2  | \langle \psi_{nl} | \bs r | \Psi_{{\bs p}_{\ma{rel}}} \rangle|^2  f_{nl}(\bs x, \bs k ,t=0) \,.
\ee
The other term in the anti-commutator gives the same result. So applying the Wigner transform to the $-\frac{1}{2} \gamma_{ab,cd} \langle {\bs k}_1, n_1l_1, 1 |\{L^{\dagger}_{cd}L_{ab}, \rho_S(0)\} | {\bs k}_2, n_2l_2, 1\rangle$ term in the Lindblad equation,
yields the negative of Eq.~(\ref{eqn:disso_boltz}).

\section{$\gamma_{ab,cd}L_{ab}\rho_S(0)L_{cd}^{\dagger}$ term}
\label{app:recombination}
For this term we set $|a\rangle = |{\bs k}_1, n_1l_1,1 \rangle$, $|c\rangle=|{\bs k}_2, n_2l_2,1\rangle$, $|b\rangle=|{\bs p}_{1\ma{cm}}, {\bs p}_{1\ma{rel}},a_1\rangle$ and $|d\rangle=|{\bs p}_{2\ma{cm}}, {\bs p}_{2\ma{rel}},a_2\rangle$ where $a_1$ and $a_2$ are color indexes. We need to evaluate
\be \nn
\gamma_{ab,cd} &=& \int \diff^3R_1 \int \diff^3R_2 \sum_{i_1,i_2,b_1,b_2}\int_0^t \diff t_1 \int_0^t \diff t_2 C_{{\bs R}_2i_2b_2,{\bs R}_1i_1b_1}(t_2, t_1) \\ \nn
&&\langle {\bs k}_1, n_1l_1, 1 | \langle S({\bs R}_1, t_1) | r_{i_1} | O^{b_1}({\bs R}_1, t_1) \rangle  | {\bs p}_{1\ma{cm}}, {\bs p}_{1\ma{rel}},a_1\rangle \\
&&\langle {\bs p}_{2\ma{cm}}, {\bs p}_{2\ma{rel}},a_2 |  \langle O^{b_2}({\bs R}_2, t_2) | r_{i_2} | S({\bs R}_2, t_2) \rangle | {\bs k}_2, n_2l_2, 1\rangle \,.
%&&\langle {\bs k}_1, n_1l_1, 1 | S^{\dagger}({\bs R}_1, t_1) r_{i_1} O^{b_1}({\bs R}_1, t_1)  | {\bs p}_{1\ma{cm}}, {\bs p}_{1\ma{rel}},a_1\rangle
%\langle {\bs p}_{2\ma{cm}}, {\bs p}_{2\ma{rel}},a_2 |  O^{b_2\dagger}({\bs R}_2, t_2) r_{i_2} S({\bs R}_2, t_2) | {\bs k}_2, n_2l_2, 1\rangle \,.
\ee
We first compute the singlet-octet transition term,
\be \nn
&&  \langle {\bs k}_1, n_1l_1, 1 | \langle S({\bs R}_1, t_1) | r_{i_1} | O^{b_1}({\bs R}_1, t_1) \rangle  | {\bs p}_{1\ma{cm}}, {\bs p}_{1\ma{rel}},a_1\rangle \\\nn
%&&\langle {\bs k}_1, n_1l_1, 1 | S^{\dagger}({\bs R}_1, t_1) r_{i_1} O^{b_1}({\bs R}_1, t_1)  | {\bs p}_{1\ma{cm}}, {\bs p}_{1\ma{rel}},a_1\rangle \\\nn
  &=&  \langle \psi_{n_1l_1} | r_{i_1}| \Psi_{{\bs p}_{1\ma{rel}}} \rangle \delta^{a_1b_1}e^{-i(E_{{\bs p}_1}t_1 - {\bs p}_{1\ma{cm}} \cdot {\bs R}_1)} e^{i(E_{{\bs k}_1}t_1 - {\bs k}_1\cdot{\bs R}_1)} \\
 &&\langle {\bs p}_{2\ma{cm}}, {\bs p}_{2\ma{rel}},a_2 |  \langle O^{b_2}({\bs R}_2, t_2) | r_{i_2} | S({\bs R}_2, t_2) \rangle | {\bs k}_2, n_2l_2, 1\rangle \\\nn
%&&\langle {\bs p}_{2\ma{cm}}, {\bs p}_{2\ma{rel}},a_2 |  O^{b_2\dagger}({\bs R}_2, t_2) r_{i_2} S({\bs R}_2, t_2) | {\bs k}_2, n_2l_2, 1\rangle \\\nn
 &=&  \langle \Psi_{{\bs p}_{2\ma{rel}}}   | r_{i_2}|  \psi_{n_2l_2} \rangle \delta^{a_2b_2}e^{- i(E_{{\bs k}_2}t_2 - {\bs k}_2\cdot{\bs R}_2 )} e^{i(E_{{\bs p}_2}t_2 - {\bs p}_{2\ma{cm}} \cdot {\bs R}_2)} 
\ee
The correlation in real-time thermal field theory is
\be \nn
&&C_{{\bs R}_2i_2b_2,{\bs R}_1i_1b_1}(t_2, t_1) = \frac{T_F}{N_C}g^2\langle  E_{i_2}^{b_2}({\bs R}_2, t_2)   E_{i_1}^{b_1}({\bs R}_1, t_1) \rangle_T \\ \nn
&=&\frac{T_F}{N_C}g^2 \delta^{b_1b_2} \int\frac{\diff^4q}{(2\pi)^4}  e^{iq_0(t_1-t_2)-i\bs q\cdot({\bs R}_1-{\bs R}_2)} \\
&&(q_0^2\delta_{i_1i_2} -q_{i_1}q_{i_2}) (1+n_B(q_0))   (2\pi) \sign(q_0) \delta(q_0^2-{\bs q}^2) +\ml{O}(g^3)\,,
\ee
where we have used the expression of $D^{>\,ab}_{\mu\nu}(q)$ in Appendix~\ref{app:real-time}.

Now we can combine everything and write the $  \langle {\bs k}_1, n_1l_1, 1 | \gamma_{ab,cd}L_{ab}\rho_S(0)L_{cd}^{\dagger} | {\bs k}_2, n_2l_2, 1\rangle $ term out explicitly
\be\nn
&&   \int\frac{\diff^4q}{(2\pi)^4}\frac{\diff^3 p_{1\ma{cm}}}{(2\pi)^3} \frac{\diff^3 p_{1\ma{rel}}}{(2\pi)^3} \frac{\diff^3 p_{2\ma{cm}}}{(2\pi)^3} \frac{\diff^3 p_{2\ma{rel}}}{(2\pi)^3}  \int \diff^3R_1 \int \diff^3R_2 \int_0^t \diff t_1 \int_0^t \diff t_2  \sum_{a_1,a_2, b_1,b_2, i_1,i_2} \\ \nn
&&   \frac{T_F}{N_C}g^2 \delta^{b_1b_2}    (q_0^2\delta_{i_1i_2}-q_{i_1}q_{i_2})(1+n_B(q_0)) (2\pi)\sign(q_0)\delta(q_0^2-{\bs q}^2)   e^{iq_0(t_1-t_2)-i\bs q\cdot({\bs R}_1-{\bs R}_2)}  \\ \nn
&&   \langle \psi_{n_1l_1} | r_{i_1}| \Psi_{{\bs p}_{1\ma{rel}}} \rangle \delta^{a_1b_1} e^{-i(E_{{\bs p}_1}t_1 - {\bs p}_{1\ma{cm}} \cdot {\bs R}_1)} e^{i(E_{{\bs k}_1}t_1 - {\bs k}_1\cdot{\bs R}_1)} \\ \nn
&& \langle \Psi_{{\bs p}_{2\ma{rel}}}   | r_{i_2}|  \psi_{n_2l_2} \rangle \delta^{a_2b_2}  e^{-i(E_{{\bs k}_2}t_2 - {\bs k}_2\cdot{\bs R}_2 )} e^{i(E_{{\bs p}_2}t_2 - {\bs p}_{2\ma{cm}} \cdot {\bs R}_2)} \\ 
&&\langle {\bs p}_{1\ma{cm}}, {\bs p}_{1\ma{rel}},a_1  |\rho_S(0) | {\bs p}_{2\ma{cm}}, {\bs p}_{2\ma{rel}},a_2 \rangle \,.
\ee
Integrating over ${\bs R}_1$ and ${\bs R}_2$ gives two delta functions in momenta $\delta^3({\bs k}_1-{\bs p}_{1\ma{cm}}+\bs q)\delta^3({\bs k}_2-{\bs p}_{2\ma{cm}}+\bs q)$. Under the Markovian approximation, $t\rightarrow\infty$, integrating over $t_1$ and $t_2$ gives another two delta functions $\delta(E_{k_1}-E_{p_1}+q_0)\delta(E_{k_2}-E_{p_2}+q_0)$. Then $q_0$ has to be positive because $E_{k_i}<0$ and $E_{p_i}>0$. (Using the representation $D^{<\,ab}_{\mu\nu}(q)$ will give $q_0<0$ but lead to the same result due to $1+n_B(q_0)+n_B(-q_0) = 0$). Again we set $q_0^2\delta_{i_1i_2}-q_{i_1}q_{i_2} \rightarrow \frac{2}{3}q^2\delta_{i_1i_2}$ since the gluon is on shell $q_0 = |\bs q| =q $. Now we have
\be \nn
\label{eqn:reco}
&&\int \frac{\diff^3 p_{1\ma{cm}}}{(2\pi)^3} \frac{\diff^3 p_{1\ma{rel}}}{(2\pi)^3} \frac{\diff^3 p_{2\ma{cm}}}{(2\pi)^3} \frac{\diff^3 p_{2\ma{rel}}}{(2\pi)^3}  \frac{\diff^3q}{(2\pi)^32q} (1+n_B(q)) \sum_{a, i}  \\ \nn
&&(2\pi)^8\delta^3({\bs k}_1-{\bs p}_{1\ma{cm}}+\bs q)\delta^3({\bs k}_2-{\bs p}_{2\ma{cm}}+\bs q)\delta(E_{k_1}-E_{p_1}+q)\delta(E_{k_2}-E_{p_2}+q) \\\nn
&& \frac{2T_F}{3N_C}q^2g^2 \langle \psi_{n_1l_1} | r_{i}| \Psi_{{\bs p}_{1\ma{rel}}} \rangle\langle \Psi_{{\bs p}_{2\ma{rel}}}   | r_{i}|  \psi_{n_2l_2} \rangle \\
&&\langle {\bs p}_{1\ma{cm}}, {\bs p}_{1\ma{rel}},a  |\rho_S(0) | {\bs p}_{2\ma{cm}}, {\bs p}_{2\ma{rel}},a \rangle \,.
\ee
Before integrating the $\delta$-functions, we first apply the Wigner transform on Eq.~(\ref{eqn:reco}) (by setting ${\bs k}_1 = {\bs k} + \frac{{\bs k}'}{2}$, ${\bs k}_2 = {\bs k}-\frac{{\bs k}'}{2}$,  $n_1=n_2=n$ and $l_1=l_2=l$):
\be \nn
&& \int\frac{\diff^3k'}{(2\pi)^3} e^{i {\bs k}'\cdot {\bs x} } 
\frac{\diff^3 p_{1\ma{cm}}}{(2\pi)^3} \frac{\diff^3 p_{1\ma{rel}}}{(2\pi)^3} \frac{\diff^3 p_{2\ma{cm}}}{(2\pi)^3} \frac{\diff^3 p_{2\ma{rel}}}{(2\pi)^3}  \frac{\diff^3q}{(2\pi)^32q} (1+n_B(q)) \sum_{a, i}  \\ \nn
&&(2\pi)^8\delta^3({\bs k}+\frac{{\bs k}'}{2}-{\bs p}_{1\ma{cm}}+\bs q)\delta^3({\bs k}-\frac{{\bs k}'}{2}-{\bs p}_{2\ma{cm}}+\bs q)\delta(E_{k_1}-E_{p_1}+q)\delta(E_{k_2}-E_{p_2}+q) \\ \nn
&& \frac{2T_F}{3N_C}q^2g^2 \langle \psi_{nl} | r_{i}| \Psi_{{\bs p}_{1\ma{rel}}} \rangle\langle \Psi_{{\bs p}_{2\ma{rel}}}   | r_{i}|  \psi_{nl} \rangle \\
&& \langle {\bs p}_{1\ma{cm}}, {\bs p}_{1\ma{rel}},a  |\rho_S(0) | {\bs p}_{2\ma{cm}}, {\bs p}_{2\ma{rel}},a \rangle \,.
\ee
At order $Mv^2$, the c.m.~momentum does not enter the energy: $E_{k_i} = -|E_{nl}|$ and  $E_{p_i} =\frac{{\bs p}^2_{i\,\ma{rel}}}{M}$. If we shift the momentum
\be \nn
{\bs p}_{1\ma{cm}} &=& {\bs p}'_{1\ma{cm}} + \frac{{\bs k}'}{2} \\
{\bs p}_{2\ma{cm}} &=& {\bs p}'_{2\ma{cm}} -  \frac{{\bs k}'}{2} \,,
\ee
then the two momentum $\delta$-functions become $\delta^3({\bs k}-{\bs p}'_{1\ma{cm}}+\bs q)\delta^3({\bs k}-{\bs p}'_{2\ma{cm}}+\bs q)$. So we can integrate over ${\bs p}'_{2\ma{cm}}$ and set ${\bs p}'_{2\ma{cm}}={\bs p}'_{1\ma{cm}} = {\bs p}_{\ma{cm}}$. Due to the two energy $\delta$-functions, we have $p_{1\ma{rel}} = p_{2\ma{rel}} $. To simplify further, we assume the octet scattering wave function can be factorized,
\be
\langle \bs r | \Psi_{{\bs p}_{\ma{rel}}} \rangle = e^{i   {\bs p}_{\ma{rel}}   \cdot   \bs r} f(r, p_{\ma{rel}})\,,
\ee
which is true for the plane wave solution. 
If we further let
\be \nn
{\bs p}_{1\ma{rel}} &=& {\bs p}_{\ma{rel}} \\
{\bs p}_{2\ma{rel}} &=& {\bs p}_{\ma{rel}} + {\bs p}'_{\ma{rel}} \,,
\ee
(remember that we have shown $p_{1\ma{rel}}=p_{2\ma{rel}}$,) we obtain
\be\nn
&& t\int\frac{\diff^3 p_{\ma{cm}}}{(2\pi)^3} \frac{\diff^3 p_{\ma{rel}}}{(2\pi)^3} \frac{\diff^3q}{(2\pi)^32q} (1+n_B(q)) \sum_{a,i}
(2\pi)^4\delta^3({\bs k}-{\bs p}_{\ma{cm}}+\bs q) \delta(-|E_{nl}|+q-\frac{{\bs p}_{\ma{rel}}^2}{M})\\ \nn
&& \frac{2T_F}{3N_C}q^2g^2 \langle  \psi_{nl}  | r_i |  \Psi_{{\bs p}_{\ma{rel}}}  \rangle \int \diff^3r \,  \psi_{nl}(\bs r) r_i \Psi^*_{{\bs p}_{\ma{rel}}}(\bs r) \\
\label{eqn:reco_contribute}
&& \int \frac{\diff^3p'_{\ma{rel}}}{(2\pi)^3} e^{-i  {\bs p}'_{\ma{rel}}  \cdot  \bs r } \int \frac{\diff^3k'}{(2\pi)^3} e^{i {\bs k}'\cdot {\bs x} }  
 \langle {\bs p}_{\ma{cm}}+\frac{{\bs k}'}{2}, {\bs p}_{\ma{rel}},a  |\rho_S(0) | {\bs p}_{\ma{cm}}-\frac{{\bs k}'}{2}, {\bs p}_{\ma{rel}}+{\bs p}'_{\ma{rel}},a \rangle \,.
\ee
The last line is just the phase space distribution function of a heavy quark antiquark pair whose c.m.~position is located at $\bs x$ and whose relative position is $\bs r$:
\be \nn
 f_{Q\bar{Q}}(\bs x, {\bs p}_{\ma{cm}}, {\bs r}, {\bs p}_{\ma{rel}}, a,t=0) &=& \int \frac{\diff^3k'}{(2\pi)^3} e^{i {\bs k}'\cdot {\bs x} } \int \frac{\diff^3p'_{\ma{rel}}}{(2\pi)^3} e^{-i  {\bs p}'_{\ma{rel}}    \cdot   \bs r } \\
&& \langle {\bs p}_{\ma{cm}}+\frac{{\bs k}'}{2}, {\bs p}_{\ma{rel}}, a  |\rho_S(0) | {\bs p}_{\ma{cm}}-\frac{{\bs k}'}{2}, {\bs p}_{\ma{rel}} + {\bs p}'_{\ma{rel}},a \rangle \,.
\ee
So the $ \langle {\bs k}_1, n_1l_1, 1 |  \gamma_{ab,cd}L_{ab}\rho_S(0)L_{cd}^{\dagger}  | {\bs k}_2, n_2l_2, 1\rangle$ term in the Lindblad equation under a Wigner transform leads to
\be\nn
&& t\int\frac{\diff^3 p_{\ma{cm}}}{(2\pi)^3} \frac{\diff^3 p_{\ma{rel}}}{(2\pi)^3} \frac{\diff^3q}{(2\pi)^32q} (1+n_B(q))  
 (2\pi)^4\delta^3({\bs k}-{\bs p}_{\ma{cm}}+\bs q) \delta(-|E_{nl}|+q-\frac{{\bs p}_{\ma{rel}}^2}{M})\\ 
&& \sum_{a,i} \frac{2T_F}{3N_C}q^2g^2 \langle  \psi_{nl}  | r_i |  \Psi_{{\bs p}_{\ma{rel}}}  \rangle \int \diff^3r \,  \psi_{nl}(\bs r) r_i \Psi^*_{{\bs p}_{\ma{rel}}}(\bs r) 
 f_{Q\bar{Q}}(\bs x, {\bs p}_{\ma{cm}}, {\bs r}, {\bs p}_{\ma{rel}}, a,t=0) \,.
\ee

\bibliographystyle{apsrev4-1}

\end{document}